\documentclass[11pt,prd,aps,showpacs,nofootinbib,oneside,hidelinks,a4paper,english]{revtex4-2}

\usepackage{verbatim}
\usepackage[pdftex]{pict2e}
\usepackage[dvipsnames]{xcolor}
\usepackage[compat=1.1.0]{tikz-feynman}
\usetikzlibrary{shapes.geometric,calc,positioning}
\usepackage{tikz}
\usetikzlibrary{calc, positioning}
\usepackage{pgfplots}
\usepackage{feynmp-auto}
\usepackage{graphicx}
\usepackage{amssymb}
\usepackage{amsmath}
\usepackage[utf8]{inputenc}
\usepackage{textcomp}
\usepackage{graphics}
\usepackage{dcolumn}
\usepackage{bm}
\usepackage{color}
\usepackage{epstopdf}
\usepackage{lipsum}
\usepackage{ulem}
\usepackage{braket}
\usepackage{hyperref}
\usepackage{tikz-feynman}
\hypersetup{
	colorlinks=true,
	linkcolor=red,  % Cor das referências internas (ex: \ref, \eqref)
	citecolor=blue,  % Cor das citações bibliográficas (ex: \cite)
	urlcolor=blue   % Cor dos links da internet (ex: \url ou \href)
}

\DeclareMathOperator{\arccot}{arccot}

\pgfplotsset{compat=1.18}

\begin{document}
	
	\newcommand{\red}[1]{\textcolor{red}{#1}}
	\newcommand{\sdc}{\red}
	\newcommand{\blue}[1]{\textcolor{blue}{#1}}
	\newcommand{\rjm}{\blue}
	
	\title{Dark-State Interference in an Effective Confined Three-Level Color System}
	
	\author{S. D. Campos\footnote{sergiodc@ufscar.br}}
	\affiliation{Applied Mathematics Laboratory-DFQM/CCTS,
		Federal University of S\~ao Carlos, Sorocaba CEP 18052780, Brazil}
	
	\begin{abstract}
		In this work, a QCD-inspired effective model describes dark-state interference in a confined three-state system with two low-energy $Q\overline{Q}$ states, $|1\rangle$ and $|2\rangle$, and a gluon-excited hybrid state, $|H\rangle$. A stationary chromoelectric flux-tube background enters the unperturbed Hamiltonian, while a time-dependent chromoelectric perturbation couples $|1\rangle$ and $|2\rangle$ to $|H\rangle$ in a $\Lambda$-type configuration. Under two-photon\footnote{We retain standard terms like ``two-photon'' from the EIT literature only by mathematical analogy, but they do not refer to electromagnetic photons but to the effective frequencies and energy mismatches of chromoelectric perturbations and gluonic excitations that connect confined states.} resonance and within a Markovian Gorini–Kossakowski–Sudarshan-Lindblad master equation, the system forms a coherent superposition that decouples from $|H\rangle$, strongly suppressing the steady-state probe susceptibility $\operatorname{Im}\chi_{\text{CE}}$ as lower-state decoherence vanishes. A phenomenological non-Abelian correction to the hybrid energy mainly shifts and reshapes the transparency window rather than creating the interference itself. The model is not a first-principles QCD calculation of medium opacity but a controlled effective link between dark-state interference in quantum optics and reduced confined color dynamics.
	\end{abstract}
	
	\date{\today}
	\maketitle
	
	%%%%%%%%%%%%%%%%%%%%%%%%%%%%%%%%%%
	\section{Introduction}\label{sec:intro}
	
	The current understanding identifies Quantum Chromodynamics (QCD) as the fundamental theory of the strong interaction across the full energy range accessible to current experiments~\cite{PDG.2022}. Its predicted elementary degrees of freedom, quarks ($Q$) and gluons ($G$), have not yet been observed as free particles in nature, and instead appear only as colorless composite states known as hadrons~\cite{M.E.Peskin.D.V.Schroeder.book.1995}. The hadron itself represents color confinement \cite{grosswilczek.1973,politzer.1973} and is one of the most distinctive and challenging features of the strong interaction. 
	
	Nevertheless, at high temperature and small baryochemical potential \(\mu_B\), the confined phase undergoes a transition to deconfined quark and gluon matter, known as the quark–gluon plasma (QGP). It is produced transiently in ultra-relativistic heavy-ion collisions at CERN's Super Proton Synchrotron (SPS) \cite{alber1995,heinz2000} and BNL's Relativistic Heavy Ion Collider (RHIC) \cite{adams2005,back2005,arsene2005}. The QGP lifetime~\cite{buszarajagopal.2018}, about $10\ \mathrm{fm}/c \sim 10^{-22} \mathrm{s}$, precludes direct detection of free $Q$ and $G$; thus, its properties must be inferred indirectly from hadronic and electromagnetic probes emitted during the collision.
	
	%At sufficiently high temperatures and vanishing chemical potential, QCD predicts a transition to a deconfined state of $Q$ and $G$, the QGP, which is produced in ultra-relativistic heavy-ion collisions at CERN's Super Proton Synchrotron (SPS)~\cite{alber1995,heinz2000} and BNL's Relativistic Heavy Ion Collider (RHIC)~\cite{adams2005,back2005,arsene2005}. The QGP lifetime~\cite{buszarajagopal.2018}, on the order of $10~\mathrm{fm}/c \approx 10^{-22}\,\mathrm{s}$, precludes direct observation of deconfined $Q$ and $G$; hence, its properties must be inferred indirectly from hadronic and electromagnetic (EM) probes emitted during the collision.
	
	Of course, the theoretical interpretation of these probes requires a detailed understanding of how interactions between color charges evolve across different energy scales and phases. In the perturbative regime of QCD, $QG$ interactions are well approximated by a Coulomb-like potential, while in the confinement regime, dynamics are governed by strongly non-perturbative chromoelectric (CE) and chromomagnetic fields, commonly modeled as color flux tubes connecting a quark and antiquark ($Q\bar{Q}$) pair~\cite{H.Fritzsch.M.Gell-Mann.1972,bagan1991}. Recent lattice studies of pure $SU(3)$ gauge theory show that tube-like CE field profiles between static sources persist well above the deconfinement critical temperature $T_c$, at least up to $T \simeq 1.5\,T_c$~\cite{shuryak.2018,liao.2007}. This represents a non-perturbative regime, where the gluon mediates the interaction while carrying a color charge and self-interacting~\cite{H.Fritzsch.M.Gell-Mann.1972}. The transverse profile remains roughly flux-tube-like and even narrows as temperature rises, rather than returning to a near-spherical Coulomb behavior~\cite{shuryak.2018,liao.2007}. Even this setup, a single $Q\bar{Q}$ pair connected by a flux tube, already requires technically demanding calculations~\cite{bali.1995,cea.2012} and provides a natural setting for exploring analogies between QCD and other coherent quantum systems~\cite{mandelstam.1976,kharzeev.2017}, as done here.
	
	A paradigmatic example from quantum optics is electromagnetically induced transparency (EIT), which arises in driven three-level systems with a strong coupling field and a weak probe field~\cite{S.E.Harris.J.E.Field.A.Imamoglu.Phys.Rev.Lett.64.10.1107.1990,ElectromagneticallyInducedTransparency.RevModPhys.77.633.2005}. From a theoretical perspective, EIT is often described using three-level Hamiltonian models, density-matrix coherences, and the resulting optical susceptibility of the medium~\cite{ElectromagneticallyInducedTransparency.RevModPhys.77.633.2005}. EIT has interesting analogies with other physical systems, such as EIT-like behavior in continuous media, including plasmon-induced transparency~\cite{zhang.2008} and fast quantum gas formation~\cite{xin.2025}. Under suitable resonance conditions, destructive quantum interference among multiple excitation pathways significantly reduces probe-field absorption within a narrow transparency window. This phenomenon is associated with coherent population trapping and non-absorbing (dark) superposition states, and it has been extensively studied both experimentally and theoretically in quantum-mechanical and effectively classical systems (see, for example, Refs. \cite{G.Alzetta.A.Gozzini.L.Moi.G.Orriols.Nuovo.Cimento.B.36.5.1976,C.L.G.Alzar.M.A.G.Martinez.P.Nussenzveig.Am.J.Phys.70.37.2002}). 
	
	Considering the above paragraphs, the main question here is conceptual: 
	{\it can the minimal algebraic structure underlying EIT, destructive interference between coherent transition amplitudes in an effective three-level system, be realized in a confined color sector?} More specifically, 
	{\it can a reduced QCD-inspired Hamiltonian support a dark-state mechanism analogous to a $\Lambda$-type optical system?} A follow-up question is: 
	{\it if such a mechanism exists, how does a phenomenological non-Abelian contribution modify the resulting transparency-like spectral feature?}  
	
	To address these questions, one introduces a reduced three-state description of a confined heavy $Q\bar Q$ system with three effective states: two low-lying $Q\bar Q$ configurations, $\lvert 1\rangle$ and $\lvert 2\rangle$, and a gluon-excited hybrid configuration $\lvert H\rangle$. In this truncated Hilbert space, $\lvert 1\rangle$ and $\lvert 2\rangle$ are the two lower states of a $\Lambda$-type system, and $\lvert H\rangle$ is the common hybrid channel. A time-dependent CE perturbation drives coherent transitions between the lower states and $\lvert H\rangle$. Under suitable resonance and coherence conditions, the transition amplitudes interfere destructively, producing a superposition that decouples from $\lvert H\rangle$. This framework should be understood as an effective few-level model of confined color configurations, not a literal one-to-one mapping between individual $Q$ and $G$ degrees of freedom and atomic energy levels. 
	
	We formulate this EIT-like mechanism in a reduced QCD-inspired Hamiltonian framework and examine how a phenomenological non-Abelian correction modifies the resulting spectral response. The model also identifies which predictions require microscopic input from lattice QCD or Born–Oppenheimer calculations.
	
	%The main purpose here is to test whether the dark-state mechanism characteristic of EIT can be consistently formulated within a reduced, QCD-inspired Hamiltonian framework. Specifically, the interference mechanism is treated as an intrinsic structural feature of the three-level coupling scheme, while the non-Abelian contribution is incorporated phenomenologically as a correction to the hybrid-state energy. The main purpose is to show that the minimal algebraic mechanism behind EIT, destructive interference between two coherent transition pathways, can be formulated consistently within a reduced confined-state description. This result can clarify how a diagonal non-Abelian correction alters the spectral position and shape of the resulting transparency-like feature. Of course, this mechanism can provide control scenarios and dimensionless reference cases, showing how the transparency window depends on lower-state decoherence, coupling strengths, and flux-tube geometry, and highlighting which aspects require microscopic input (e.g., from lattice or Born–Oppenheimer calculations) for quantitative predictions.
	
	This work is organized as follows. Section~\ref{sec:chro_elec_int} defines the reduced confined-state Hilbert space, distinguishes the stationary confining background from the time-dependent driving perturbation, and introduces the effective three-level Hamiltonian, from which the CE dark-state and conditions for destructive interference are derived. Section~\ref{sec:opacity_control} derives the induced CE response from the density-matrix dynamics and defines an EIT-like susceptibility. Section~\ref{sec:control_cases} presents control cases that separate the three-level interference mechanism from the non-Abelian hybrid-energy correction. Theoretical predictions are presented in Section~\ref{sec:theo_pred}. Section~\ref{sec:time} presents a discussion of the predicted lifetime for this phenomenon. Finally, Section~\ref{sec:final_remarks} summarizes the scope and limitations of the effective construction. %Throughout, the distinction is kept clear between the generic dark-state interference mechanism and the specifically non-Abelian contribution that shifts or reshapes the corresponding transparency feature.
	
	%%%%%%%%%%%%%%%%%%%%%%%%%%%%%%%%%%%%%%%%%%%%%%%%%%%%%%%%%%%%%%%%%%%%%%%%%
	\section{CE field interaction}
	\label{sec:chro_elec_int}
	
	As is well known, in an Abelian gauge theory, the electric field can be expressed in terms of electromagnetic potentials. In QCD, by contrast, the CE field includes an additional nonlinear contribution associated with the non-Abelian gauge structure \cite{M.E.Peskin.D.V.Schroeder.book.1995}. In terms of the gluon gauge potential $\mathcal{A}_{\mu}^{a}$, the CE field is given by \cite{M.E.Peskin.D.V.Schroeder.book.1995}
	\begin{equation}
		\mathcal{E}_{i}^{a} = \partial_{i}\mathcal{A}_{0}^{a} - \partial_{0}\mathcal{A}_{i}^{a} + g f^{abc}\mathcal{A}_{i}^{b}\mathcal{A}_{0}^{c},
		\label{eq:nonabelian_ce_field}
	\end{equation}
	where $g$ is the QCD gauge coupling, $a,b,c$ are adjoint color indices, and $f^{abc}$ are the structure constants of the gauge group. The last term on the right-hand side has no electromagnetic counterpart and comes from gluon self-interaction. In this work, equation~\eqref{eq:nonabelian_ce_field} mainly motivates a non-Abelian correction to the reduced model, rather than representing a first-principles solution of the full Yang–Mills dynamics.
	
	%%%%%%%%%%%%%%%%%%%%%%%%%%%%%%%%%%%%%%%55
	%\subsection{Reduced confined-state Hilbert space}
	%\label{subsec:reduced_hilbert}
	
	Consider a heavy $Q\bar Q$ pair at fixed separation $r$, with gluonic degrees of freedom restricted to a small set of effective confined configurations. The full QCD Hilbert space is projected onto the three-dimensional subspace
	\begin{equation}
		\mathcal{H}_{\mathrm{eff}} = \operatorname{span}
		\left\{\lvert 1\rangle,\lvert 2\rangle,\lvert H\rangle\right\},
		\label{eq:effective_hilbert_space}
	\end{equation}
	where the states $\lvert 1\rangle$ and $\lvert 2\rangle$ denote two orthonormal low-energy configurations of the confined $Q\bar{Q}$ sector, and $\lvert H\rangle$ denotes the hybrid configuration in which the gluonic field is excited. They form an orthonormal set satisfying the condition
	\begin{equation}
		\langle i \vert j\rangle = \delta_{ij},
		\qquad
		i,j\in\{1,2,H\}.
		\label{eq:orthonormal_effective_states}
	\end{equation}
	
	Of course, the basis vectors in subspace~\eqref{eq:effective_hilbert_space} do not correspond to isolated, colored asymptotic particle states: each vector describes the full confined system, including both the heavy color sources and the mediating gluonic field between them. As a result, this three-state formulation is a truncation of the full confined spectrum and is assumed to provide an adequate description only within the limited energy and time scales considered in this work (as shall be seen later).
	
	At the microscopic level, gauge-covariant $Q\bar{Q}$ configurations can be constructed by dressing the heavy color sources with Wilson lines \cite{lavelle.1997,brambilla.2005}, and a general form of a confined configuration can be expressed as \cite{M.E.Peskin.D.V.Schroeder.book.1995,brambilla.2000}
	\begin{equation}
		\lvert Q\bar{Q};n\rangle \sim \bar{Q}(\mathbf{x}_{2})
		W(\mathbf{x}_{2},\mathbf{x}_{1}) \Gamma_{n} Q(\mathbf{x}_{1}) \lvert 0\rangle,
		\label{eq:wilson_confined_state}
	\end{equation}
	where
	\begin{equation}
		W(\mathbf{x}_{2},\mathbf{x}_{1}) = \mathcal{P} \exp
		\left[ig\int_{\mathbf{x}_{1}}^{\mathbf{x}_{2}} \mathcal{A}_{i}^{a}(\mathbf{z})T^{a}\,dz^{i} \right]
		\label{eq:wilson_line}
	\end{equation}
	is a Wilson line connecting the $Q$ and $\bar{Q}$ positions \cite{brambilla.2000,wilson.1974}. 
	
	In contrast to conventional mesonic states, hybrid configurations include explicit gluonic degrees of freedom that contribute to the system's quantum numbers \cite{dudek.2010,meyer2015}. Microscopically, such a state can be realized by inserting a local gluonic field operator $\mathcal{O}_g(\mathbf{z})$, such as the CE field $\mathcal{E}^a(r,s)$ or the chromomagnetic field along the color flux tube connecting the heavy quark–antiquark pair \cite{brambilla.2000,dudek.2010}. To preserve local $SU(3)$ gauge invariance, the original gauge link is split into two Wilson lines $W(\mathbf{x}_2,\mathbf{z})$ and $W(\mathbf{z},\mathbf{x}_1)$ that transform covariantly with $\mathcal{O}_g(\mathbf{z})$ under local gauge transformations \cite{dudek.2010}, resulting in the hybrid state given by
	\begin{equation}
		\lvert H\rangle \sim \bar{Q}(\mathbf{x}_{2})
		W(\mathbf{x}_{2},\mathbf{z}) \mathcal{O}_{g}(\mathbf{z})
		W(\mathbf{z},\mathbf{x}_{1}) Q(\mathbf{x}_{1})
		\lvert 0\rangle.
		\label{eq:hybrid_state_schematic}
	\end{equation}
	
	We emphasize that this work does not aim to derive these nonperturbative configurations from first-principles QCD calculations: energies and transition matrix elements are treated as effective phenomenological parameters, serving as model inputs that encode their impact on the dynamics without a microscopic derivation. Taking into account this observation, the unperturbed Hamiltonian can be defined by
	\begin{equation}
		\hat{H}_{0}\lvert j\rangle = \omega_{j}\lvert j\rangle,
		\qquad
		j\in\{1,2,H\},
		\label{eq:unperturbed_effective_states}
	\end{equation}
	where $\omega_{1}$ and $\omega_{2}$ are the energies of the two low-lying confined configurations, and $\omega_{H}$ is the energy of the gluon-excited hybrid configuration. The general state in the reduced subspace can be written as
	\begin{equation}
		\lvert\Psi(t)\rangle = c_{1}(t)e^{-i\omega_{1}t}\lvert 1\rangle
		+ c_{2}(t)e^{-i\omega_{2}t}\lvert 2\rangle
		+ c_{H}(t)e^{-i\omega_{H}t}\lvert H\rangle,
		\label{eq:three_level_state}
	\end{equation}
	assuming the normalization $\lvert c_{1}(t)\rvert^{2} + \lvert c_{2}(t)\rvert^{2} + \lvert c_{H}(t)\rvert^{2} = 1$.

	%%%%%%%%%%%%%%%%%%%%%%%%%%%%%%%%%%%%%%%%%%%%%%%
	\subsection{Static confining background and time-dependent driving field}
	\label{subsec:background_and_drive}
	
	It is essential to distinguish the stationary CE background field, which supports the formation of the confining flux tube, from the time-dependent perturbation that induces transitions between the corresponding effective states. Thus, one considers here the total CE field given by equation \eqref{eq:nonabelian_ce_field} written as
	\begin{equation}
		\boldsymbol{\mathcal{E}}_{\mathrm{tot}}^{a}(r,s,t) = \boldsymbol{\mathcal{E}}_{\mathrm{tube}}^{a}(r,s)
		+ \delta\boldsymbol{\mathcal{E}}^{a}(r,s,t),
		\label{eq:field_decomposition}
	\end{equation}
	where $\delta\ll 1$ represents a small perturbation parameter and, consequently, the time-dependent field ${\mathcal{E}}^{a}(r,s,t)$ does not give rise to confinement. Instead, it acts as a perturbation of a confined \(Q\bar{Q}\) background and induces transitions among its effective configurations. In contrast, the static contribution $\boldsymbol{\mathcal{E}}_{\mathrm{tube}}^{a}(r,s)$ is incorporated into $\hat{H}_{0}$ and determines the energies and spatial structure of the confined configurations. 
	
	In order to generate the two coherent transition amplitudes required by the $\Lambda$-type model \cite{ElectromagneticallyInducedTransparency.RevModPhys.77.633.2005}, the time-dependent perturbation is expressed as
	\begin{equation}
		\delta\boldsymbol{\mathcal{E}}^{a}(r,s,t) =
		\boldsymbol{\mathcal{E}}_{p}^{a}(r,s)
		\cos(\omega_{p}t) + \boldsymbol{\mathcal{E}}_{c}^{a}(r,s)
		\cos(\omega_{c}t),
		\label{eq:driving_field}
	\end{equation}
	where $\boldsymbol{\mathcal{E}}_{p}^{a}(r,s)$ and $\boldsymbol{\mathcal{E}}_{c}^{a}(r,s)$ are the spatial amplitudes of the probe-like and coupling-like CE components, driven by frequencies $\omega_p$ and $\omega_c$, respectively. 
	
	In conventional EIT, $\omega_p$ is the probe laser frequency driving the transition between the ground state $\lvert 1\rangle$ and the excited state $\lvert e\rangle$ \cite{boller1991}; here, it denotes the CE perturbation frequency that couples the low-energy $Q\bar{Q}$ configuration $\lvert 1\rangle$ to the common hybrid channel $\lvert H\rangle$. Similarly, while the coupling frequency $\omega_c$ in EIT controls the transparency-window width through the Rabi frequency $\Omega_c$ \cite{harris1997}, here it determines the interference strength and spatial profile of the transparency window through the effective Rabi frequency $\Omega_{2H}(r,s)$.
	
	Observe that the probe-like component drives the \(\lvert 1\rangle\leftrightarrow\lvert H\rangle\) transition, while the coupling-like component drives the \(\lvert 2\rangle\leftrightarrow\lvert H\rangle\) transition. For the above approach, the corresponding Hamiltonian is written as
	\begin{equation}
		\hat{H}(t) = \hat{H}_{0} + \hat{V}(t),
		\label{eq:total_hamiltonian}
	\end{equation}
	where the interaction term is modeled through an effective CE field operator,
	\begin{equation}
		\hat{V}(t) = -g\,\hat{\mathbf{d}}^{\,a}_{\mathrm{eff}} \cdot
		\delta\boldsymbol{\mathcal{E}}^{a}(r,s,t),
		\label{eq:chromodipole_interaction}
	\end{equation}
	where $\hat{\mathbf{d}}^{\,a}_{\mathrm{eff}}$ is an effective CE transition operator acting within $\mathcal{H}_{\mathrm{eff}}$, and with matrix elements given by
	\begin{equation}
		\mathbf{d}_{ij}^{a} = \langle i|
		\hat{\mathbf{d}}^{\,a}_{\mathrm{eff}} |j\rangle
		\label{eq:ortho_operator}
	\end{equation}
	parameterizing the strength of CE transitions between the
	effective confined configurations. This operator represents a projection of the underlying QCD CE field interaction onto the reduced (truncated) three-state subspace, rather than an observable permanent color dipole moment. Restricting the interaction to the two transitions that connect the low-energy configurations to the hybrid channel, one has (see Appendix A)
	\begin{equation}
		\hat{V}(t) = \frac{1}{2} \left[ \Omega_{1H}(r,s)e^{-i\omega_{p}t}
		\lvert H\rangle\langle 1\rvert + \Omega_{2H}(r,s)e^{-i\omega_{c}t}
		\lvert H\rangle\langle 2\rvert + \mathrm{h.c.} \right],
		\label{eq:interaction_reduced_basis}
	\end{equation}
	where $\omega_{p}$ and $\omega_{c}$ denote the probe- and coupling-like driving frequencies, respectively. The effective CE Rabi frequencies can be written as
	\begin{align}
		\Omega_{1H}(r,s) &= -g\, \langle H \vert \hat{\mathbf{d}}^{\,a}_{\mathrm{eff}}
		\vert 1\rangle \cdot \boldsymbol{\mathcal{E}}_{p}^{a}(r,s),
		\label{eq:rabi_1H}
		\\
		\Omega_{2H}(r,s) &= -g\, \langle H \vert
		\hat{\mathbf{d}}^{\,a}_{\mathrm{eff}}\vert 2\rangle
		\cdot \boldsymbol{\mathcal{E}}_{c}^{a}(r,s).
		\label{eq:rabi_2H}
	\end{align}
	
	Of course, the introduction of two transition amplitudes does not require interpreting the basis states as individual particles \cite{fano.1961,weinberg.book}. It only requires two coherent matrix elements that couple the lower, confined configurations to the same hybrid channel. This is the structural condition for realizing a $\Lambda$-type interference mechanism \cite{ElectromagneticallyInducedTransparency.RevModPhys.77.633.2005}. 
	
	Although the static flux-tube background is introduced through an effective profile $\mathcal{F}(r,s)$ (defined in the next subsection) rather than derived from first-principles Yang–Mills dynamics, the decomposition into a quasi-stationary flux tube and additional CE perturbations remains qualitatively compatible with the magnetic scenario of the near-$T_c$ plasma. Within that framework, CE flux tubes may persist above the deconfinement temperature as configurations dynamically stabilized by scattering off thermal monopoles in a dual plasma~\cite{shuryak.2018,liao.2007}. The present treatment therefore models flux-tube geometry and probe-like perturbations as distinct effective degrees of freedom, consistent with this broader phenomenological picture \cite{chernodub2007}.
	
	%%%%%%%%%%%%%%%%%%%%%%%%%%%%%%%%%%%%%%%%%%%%%%%%%%%%%%%%%%%%%%
	\subsection{Flux-tube profile and effective transition amplitudes}
	\label{subsec:flux_profile}
	
	The spatial dependence of the projected transition matrix elements on interquark separation and transverse flux-tube structure is summarized by an effective geometric profile $\mathcal{F}(r,s)$ (see, for example, Refs. \cite{cardaci.2011,cea.2012,bali.1995,brambilla.2005}). Lattice-QCD calculations show that the confining CE field is concentrated between static color sources and has a nontrivial transverse profile, whose shape and width can depend on the interquark separation $r$ \cite{cardaci.2011,baker.2022}. Furthermore, the static energies of conventional and gluon-excited heavy quark configurations depend on the separation $r$, analogous to the Born–Oppenheimer treatment of heavy quarkonia and hybrid states \cite{berwein.2024}.
	
	The effective geometric profile $\mathcal{F}(r,s)$ should not be viewed as an additional fundamental field \cite{brambilla.2005,cea.2012}. Instead, it parametrizes the dependence of the projected CE transition matrix elements on the flux tube’s longitudinal extent and transverse spatial structure. Taking this into account, we write the relation
	\begin{equation}
		\boldsymbol{\mathcal{E}}_{k}^{a}(r,s) = \mathcal{A}_{k}^{a}\,
		\mathcal{F}(r,s),
		\qquad
		k\in\{p,c\},
		\label{eq:geometric_field_profile}
	\end{equation}
	such that the spatial Rabi frequencies read
	\begin{equation}
		\Omega_{kH}(r,s) = -g\,\mu_{kH}\,\mathcal{F}(r,s),
		\label{eq:rabi_geometric_profile}
	\end{equation}
	where $k \in \{p, c\}$ corresponds to the probe-like ($p$) and coupling-like ($c$) driven transitions, respectively, and
	\begin{equation}
		\mu_{kH} = \langle H \vert
		\hat{\mathbf{d}}^{\,a}_{\mathrm{eff}} \cdot
		\mathcal{A}_{k}^{a} \vert k\rangle
		\label{eq:effective_overlap}
	\end{equation}
	is the effective transition matrix element. This formulation avoids replacing the wave-function overlap with the bare spatial integral of the field profile \cite{brambilla.2000}. In the sum-rule-inspired parameterization adopted here, the longitudinal flux-tube contribution present in equation \eqref{eq:field_decomposition} is written as~\cite{bagan1991}
	\begin{equation}
		\mathcal{E}_{\mathrm{tube}}^a(r,s) = \frac{2r^{2}}
		{(N_{c}^{2}-1)(r^{2}+4s^{2})} \left[
		\sigma^{2} + \frac{2\pi\sigma\gamma}
		{3(r^{2}+4s^{2})} + \frac{\pi^{2}\gamma^{2}}
		{9(r^{2}+4s^{2})^{2}} \right],
		\label{eq:eforquarks_rew}
	\end{equation}
	where $N_{c}$ is the number of colors ($N_c = 3$), $\sigma$ is the string tension, and $\gamma$ controls subleading corrections. In principle, alternative descriptions, such as Gaussian or exponential profiles calibrated to lattice data, or flux-tube shapes derived from Born–Oppenheimer potentials, could be used without changing the underlying three-level interference mechanism. The result given by equation~\eqref{eq:eforquarks_rew} is used to define the integrated flux-tube profile \cite{cea.2012}
	\begin{equation}
		\mathcal{F}(r,s) = \int_{0}^{r} \mathcal{E}_{\mathrm{tube}}^a(r',s)\,dr',
		\label{eq:effective_geometric_profile}
	\end{equation}
	indicating it is a model-dependent geometric factor that enters the transition amplitudes through equation~\eqref{eq:rabi_geometric_profile} and cannot be identified with a complete interaction Hamiltonian. For the parameterization in equation~\eqref{eq:eforquarks_rew}, the integration of \eqref{eq:effective_geometric_profile} gives
	\begin{equation}
		\begin{split}
			\mathcal{F}(r,s) &= r
			\left[ \frac{2\sigma^{2}}{N_{c}^{2}-1} - \frac{\pi\gamma}
			{(N_{c}^{2}-1)(r^{2}+4s^{2})^{2}} \left(\frac{\pi\gamma}{18}
			+ \frac{\pi\gamma-96s^{2}\sigma}{144} \right) \right]
			\\
			&\quad
			+ \frac{ \pi^{2}\gamma^{2} + 96\pi s^{2}\gamma\sigma
				- 1152s^{4}\sigma^{2} }{ 288(N_{c}^{2}-1)s^{3}
			} \arccot\left(\frac{2s}{r}\right).
		\end{split}
		\label{eq:chromoe_x_rew}
	\end{equation}
	
\begin{figure}[ht]
		\centering
		\includegraphics[width=0.7\linewidth]{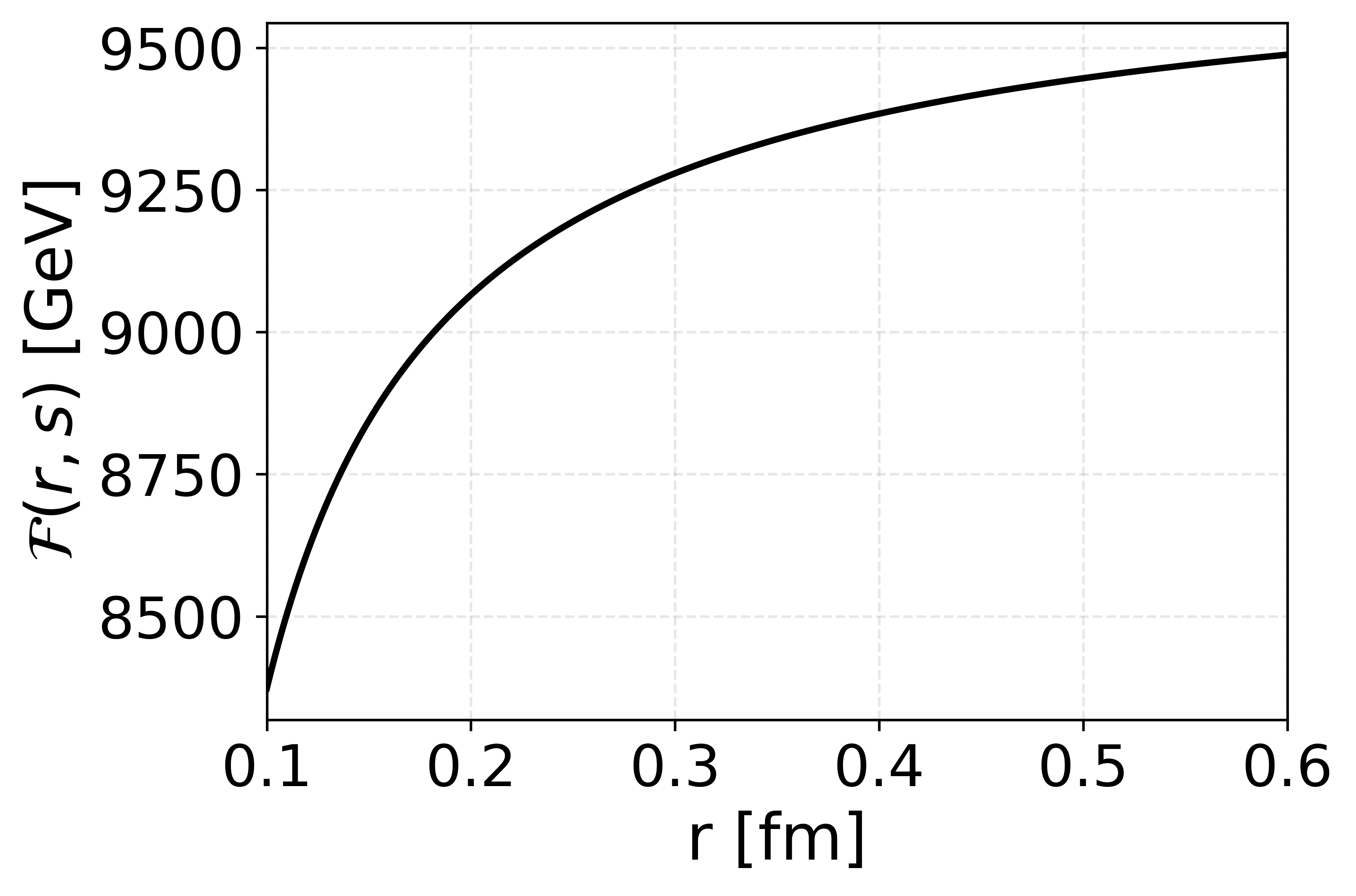}
		\caption{Effective flux-tube geometric profile $\mathcal{F}(r,s)$ versus interquark separation $r$, as defined in equation~\eqref{eq:effective_geometric_profile}. The parameters used in the sum-rule-inspired parametrization are $\sqrt{\sigma} = 0.44\text{ GeV}$, $\gamma = 1.2$, and $N_c = 3$ colors. $\mathcal{F}(r,s)$ modulates the spatial dependence of the effective CE Rabi frequencies via equation~\eqref{eq:rabi_geometric_profile} and represents an integrated geometric factor, not a complete interaction Hamiltonian or transition frequency.}
		\label{fig:fig_chromoe}
	\end{figure}

	The behavior of the integrated flux-tube profile $\mathcal{F}(r,s)$ versus interquark separation $r$ is shown in Figure~\ref{fig:fig_chromoe}. For a fixed flux-tube radius $s$, $\mathcal{F}(r,s)$ increases monotonically from the short-distance regime ($r \sim 0.1\text{ fm}$) and gradually saturates at larger separations ($r \gtrsim 0.5\text{ fm}$), reflecting the bounded longitudinal accumulation of the field profile. The numerical values and units in Figure~\ref{fig:fig_chromoe} must be interpreted according to the dimensional convention used for equation~\eqref{eq:eforquarks_rew} and the effective transition matrix elements $\mu_{kH}$.
	At this point, it is worth clarifying that the magnitude of $\mathcal{F}(r,s)$, on the order of $10^3\text{ GeV}$ as shown in Figure \ref{fig:fig_chromoe}, comes from its definition as a geometric constant integrated over the flux-tube volume. Since the physical observable (the Rabi frequency $\Omega$) depends only on $\mu_{kH} \mathcal{F}(r,s)$, the effective dipole coupling constant $\mu_{kH}$ absorbs the inverse dimensional scale, ensuring that the system's transition frequencies and energies remain within the physical scale expected for the confinement regime (sub-GeV).
	
	%%%%%%%%%%%%%%%%%%%%%%%%%%%%%%%%%%%%%%%%%%%%%%%%%%%%
	\subsection{Effective non-Abelian correction to the hybrid state}
	\label{subsec:nonabelian_hybrid_shift}
	
	The present treatment includes the gluon self-interaction only as a diagonal correction to the hybrid-state energy, neglecting off-diagonal terms that could mix the confined configurations $|1\rangle$ and $|2\rangle$ or induce additional couplings between $|H\rangle$ and higher-lying states. This implies we assume the dominant non-Abelian effect is state-dependent energy renormalization, while induced mixing within the reduced subspace can be disregarded. In a complete description, however, such off-diagonal contributions would act as additional coherent pathways that compete with or partially disrupt the interference responsible for dark-state formation, potentially preventing a perfectly dark superposition and leaving some residual absorption.
	
	This non-Abelian self-interaction yields a shifted (dressed) hybrid state energy \cite{brambilla.2000,juge1998}
	\begin{equation}
		\omega_{H}^{\mathrm{self}}(r,s) = \omega_{H}^{(0)}(r,s) + \Delta\omega_{H}^{\mathrm{NA}}(r,s),
		\label{eq:hybrid_energy_correction}
	\end{equation}
	and consequently leads to a shifted one-photon detuning,
	\begin{equation}
		\Delta_{H}^{\mathrm{self}}(r,s) = \Delta_{H}^{(0)}(r,s) - \Delta\omega_{H}^{\mathrm{NA}}(r,s).
		\label{eq:detuning_shift_control}
	\end{equation}
	where $\omega_{H}^{\mathrm{NA}}(r,s)$ accounts for the non-Abelian energy shift (or correction). Assuming a longitudinally uniform flux-tube profile, the non-Abelian energy correction depends only on the transverse radius $s$, being written as
	\begin{equation}
		\Delta\,\omega_{H}^{\mathrm{NA}}(r,s) = \kappa\,\mathcal{E}_{\perp}(s)=\kappa\left(\epsilon\frac{g}{s^{2}}\right),
		\label{eq:nonabelian_shift}
	\end{equation}
	where $\mathcal{E}_{\perp}(s)$ represents the effective transverse non-Abelian CE field intensity generated by gluon self-interactions at a transverse distance $s$ from the flux-tube axis. Moreover, $g$ is the strong coupling constant, $\epsilon$ parametrizes the relative strength of these transverse fluctuations, and $\kappa$ is an effective dimensional parameter converting the field intensity into an energy shift. From a physical standpoint, one expects $\kappa \mathcal{E}_{\perp}(s)$ to be at most comparable to the spacing between the lowest hybrid and the next gluonic excitation in the relevant Born–Oppenheimer channel, so the non-Abelian correction gives a measurable, non-dominant renormalization of $\omega_{H}^{NA}(r,s)$. In practice, reasonable estimates for $\kappa$ and $\epsilon$ should be guided by lattice determinations of hybrid static potentials and gluonic field strengths: values implying $\Delta\omega^{\mathrm{NA}}_{H}(r,s)$ much larger than known hybrid level splittings would lie outside the validity of the three-state truncation, while smaller values would correspond to a weak-shift limit in which the non-Abelian contribution has only a perturbative effect on the transparency window.
	
	%%%%%%%%%%%%%%%%%%%%%%%%%%%%%%%%%%%%%%%%%%%%%%
	\subsection{Three-level analogy and CE dark-state}
	\label{subsec:dark_state}
	
	In the rotating frame and after the rotating-wave approximation, the effective Hamiltonian in the ordered basis
	\begin{equation}
		\mathcal{B} =\left\{
		\lvert 1\rangle,
		\lvert 2\rangle,
		\lvert H\rangle
		\right\}\label{eq:ord_basis}
	\end{equation}
	is given by (see Appendix B)
	\begin{equation}
		\hat{H}_{\mathrm{eff}} = \frac{1}{2}
		\begin{pmatrix}
			0 & 0 & \Omega_{1H} \\
			0 & 2\delta & \Omega_{2H} \\
			\Omega_{1H}^{\ast} & \Omega_{2H}^{\ast} & 2\Delta_{H}
		\end{pmatrix},
		\label{eq:Heff_dark}
	\end{equation}
	where
	\begin{align}
		\Delta_{H} &= \omega_{p} - \left[\omega_{H}(r,s)-\omega_{1}\right],
		\label{eq:one_photon_detuning}
		\\
		\delta &= \left\{\omega_{p}-\left[\omega_{H}(r,s)-\omega_{1}\right]
		\right\} - \left\{\omega_{c}-\left[\omega_{H}(r,s)-\omega_{2}\right]
		\right\}
		\label{eq:two_photon_detuning}
	\end{align}
	are the one- and two-photon detunings. The non-Abelian correction in equation~\eqref{eq:hybrid_energy_correction} enters through $\omega_H(r,s)$ and therefore shifts the resonance condition.
	
	The dark-state is a coherent quantum superposition of the two lowest-energy basis states ($\vert{}1\rangle$ and $\vert{}2\rangle$) that completely cancels the probability of the system transitioning to the hybrid state ($\vert{}H\rangle$) \cite{ElectromagneticallyInducedTransparency.RevModPhys.77.633.2005}. Taking this into consideration, one writes the dark state in the present approach as
	\begin{equation}
		\lvert D_{\mathrm{CE}}\rangle = A\lvert 1\rangle
		+ B\lvert 2\rangle,
		\label{eq:dark_state_ansatz}
	\end{equation}
	where $A$ and $B$ are complex amplitudes. Assuming $\rvert A \rvert^{2} + \lvert B \rvert^{2} = 1$, it follows that
	\begin{eqnarray}
		A = \frac{\Omega_{2H}}{\sqrt{\lvert\Omega_{1H}\rvert^2 + \lvert\Omega_{2H}\rvert^2}},\quad
		B = -\frac{\Omega_{1H}}{\sqrt{\lvert\Omega_{1H}\rvert^2 + \lvert\Omega_{2H}\rvert^2}}.
	\end{eqnarray}
	
	The dark-state condition ensures the Hamiltonian has no dynamic coupling that drives the system out of the dark-state and into the excited hybrid state, being written as
	\begin{equation}
		\langle H\vert\hat{H}_{\mathrm{eff}}\vert D_{\mathrm{CE}}\rangle= 0,
		\label{eq:dark_condition}
	\end{equation}
	implying the following result
	\begin{equation}
		\Omega_{1H}^{\ast}A + \Omega_{2H}^{\ast}B = 0.
		\label{eq:dark_condition_explicit}
	\end{equation}
	
	Using the above results, one obtains the normalized dark-state given by the relation
	\begin{equation}
		\lvert D_{\mathrm{CE}}\rangle =
		\frac{\Omega_{2H}\lvert 1\rangle - \Omega_{1H}\lvert 2\rangle}{
			\sqrt{\lvert\Omega_{1H}\rvert^{2} + \lvert\Omega_{2H}\rvert^{2}}}.
		\label{eq:chromoelectric_dark_state}
	\end{equation}
	
	Observe that at two-photon resonance, $\delta\simeq 0$, implying that this state is dynamically decoupled from $\lvert H\rangle$ in the ideal coherent model. Of course, the corresponding orthogonal bright state is given by
	\begin{equation}
		\lvert B_{\mathrm{CE}}\rangle = \frac{
			\Omega_{1H}^{\ast}\lvert 1\rangle + \Omega_{2H}^{\ast}\lvert 2\rangle
		}{ \sqrt{\lvert\Omega_{1H}\rvert^{2}+\lvert\Omega_{2H}\rvert^{2}}},
		\label{eq:bright_state}
	\end{equation}
	and remains coupled to the hybrid configuration. By contrast, the dark and bright states in equations~\eqref{eq:chromoelectric_dark_state}-\eqref{eq:bright_state} are coherent superpositions within a deliberately truncated Hilbert space. For the dark state, by construction, one has the result given by equation~\eqref{eq:dark_condition_explicit}, whereas for the bright state
	\begin{equation}
		\langle H| \hat{H}_{\mathrm{eff}} |B_{\mathrm{CE}}\rangle = \frac{1}{2}
		\sqrt{|\Omega_{1H}|^{2}+|\Omega_{2H}|^{2}}.
	\end{equation}
	
	The decomposition distinguishes a superposition dynamically protected from direct hybrid excitation from one that remains coupled to the CE perturbation. The analogy with confinement is limited to dynamical accessibility: destructive interference makes the dark-state configuration inaccessible to a given transition channel, whereas confinement refers to the absence of isolated color-charged asymptotic states. No claim is made that the dark-state mechanism explains, reproduces, or provides a microscopic account of confinement.
	
	In the symmetric limit $\Omega_{1H}=\Omega_{2H}$, the dependence of $\Omega_{1H}$ and $\Omega_{2H}$ on the transverse radius $s$, interquark separation $r$, and coupling constant $g$ vanishes from the linear combination. The dark state becomes
	\begin{equation}
		\lvert D_{\mathrm{CE}}\rangle = \frac{1}{\sqrt{2}}
		\left(\lvert 1\rangle - \lvert 2\rangle \right),
		\label{eq:dark_state_symmetric}
	\end{equation}
	indicating that, in this limit, the formation of the dark state is a structural property intrinsic to the symmetry, independent of the microscopic details of the perturbation intensity. Observe that, in the present approach, suppression of the hybrid-state population results from destructive interference between two transition amplitudes that populate the same confined excitation channel. Therefore, it does not arise from the absence of CE coupling.
	
\begin{figure}[ht]
		\centering
		\begin{tikzpicture}[
			scale=1.0,
			every node/.style={font=\small},
			state/.style={circle, draw=black, thick, minimum size=12mm, inner sep=0pt},
			coupling/.style={<->, thick, >=stealth},
			shiftarrow/.style={->, >=stealth, thick, dashed, color=red!70!black}
			]
			
			% Nós dos Estados Quânticos
			\node[state, fill=black!10] (one) at (-3.6,0) {$\lvert 1\rangle$};
			\node[state, fill=black!10] (two) at (3.6,0) {$\lvert 2\rangle$};
			\node[state, fill=black!25] (hybrid) at (0,2.7) {$\lvert H\rangle$};
			
			% Conexões de Acoplamento (Acoplamento de Rabi)
			\draw[coupling] (one) -- node[above left, font=\small] {$\Omega_{1H}$} (hybrid);
			\draw[coupling] (two) -- node[above right, font=\small] {$\Omega_{2H}$} (hybrid);
			
			% Rótulos dos Estados
			\node[below=0.2cm of one, align=center]
			{low-energy confined\\configuration};
			\node[below=0.2cm of two, align=center]
			{low-energy confined\\configuration};
			
			% Rótulo do Estado Híbrido (reposicionado levemente à esquerda para dar espaço ao shift)
			\node[above left=0.3cm and -0.8cm of hybrid, align=center] (hybrid_label)
			{gluon-excited\\hybrid configuration};
			
			% Nível de Energia Deslocado (Linha Tracejada do Shift Não-Abeliano)
			%\draw[thick, dotted, color=red!70!black] ($(hybrid.north west)+(-0.3,0.6)$) -- ($(hybrid.north east)+(0.3,0.6)$);
			
			% Seta do Deslocamento de Energia Não-Abeliano (\Delta\omega_H^NA)
			\draw[shiftarrow] ($(hybrid.east)+(0.1,0)$) to[out=0, in=-90] node[right=0.1cm, align=left] 
			{$\Delta\omega_{H}^{\mathrm{NA}}(r,s) = \kappa\,\mathcal{E}_{\perp}(s)$} 
			($(hybrid.east)+(0.8,0.6)$);
			
		\end{tikzpicture}
		\caption{Schematic of the effective three-state CE system in a $\Lambda$-type configuration. States $|1\rangle$ and $|2\rangle$ are low-energy confined $Q\bar{Q}$ configurations, while $|H\rangle$ is a gluon-excited hybrid configuration. The coherent CE perturbation induces effective Rabi couplings $\Omega_{1H}$ and $\Omega_{2H}$ between the lower states and the common hybrid channel. Non-Abelian gluon self-interaction is included phenomenologically as a diagonal energy shift $\Delta\omega_H^{\text{NA}}(r,s) = \kappa\mathcal{E}_{\perp}(s)$ of the hybrid state. The diagram nodes correspond to configurations of the complete confined system, not isolated constituent particles.}
		\label{fig:effective_three_state}
	\end{figure}
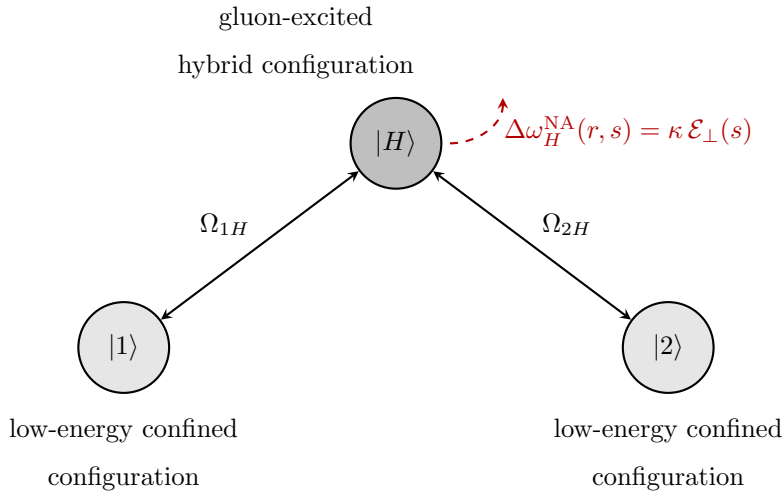

	The density matrix in the reduced basis \eqref{eq:ord_basis} is written as
	\begin{equation}
		\hat{\rho}(t) = \lvert\Psi(t)\rangle\langle\Psi(t)\rvert
		= \begin{pmatrix}
			\lvert c_{1}\rvert^{2}
			&
			c_{1}c_{2}^{\ast}
			&
			c_{1}c_{H}^{\ast}
			\\
			c_{2}c_{1}^{\ast}
			&
			\lvert c_{2}\rvert^{2}
			&
			c_{2}c_{H}^{\ast}
			\\
			c_{H}c_{1}^{\ast}
			&
			c_{H}c_{2}^{\ast}
			&
			\lvert c_{H}\rvert^{2}
		\end{pmatrix},
		\label{eq:density_matrix}
	\end{equation}
	where its off-diagonal elements describe the coherences required for the interference mechanism \cite{ElectromagneticallyInducedTransparency.RevModPhys.77.633.2005}. In particular, the cancellation in equation~\eqref{eq:dark_condition_explicit} is lost when the coherence between $\lvert 1\rangle$ and $\lvert 2\rangle$ is destroyed. Figure~\ref{fig:effective_three_state} shows the schematic interpretation of the reduced system, and the nodes correspond to configurations of the full confined system, not isolated constituent particles.
	
	Finally, truncating the full confined Hilbert space to the three-state subspace $\mathcal{H}_{\mathrm{eff}}$ \eqref{eq:effective_hilbert_space} depends on a hierarchy of energy scales. We stress that $\vert{}1\rangle$ and $\vert{}2\rangle$ denote the two lowest confined configurations in the chosen quantum-number channel, while $\vert{}H\rangle$ denotes the lowest gluon-excited hybrid state that couples efficiently to them. All higher-lying states, including radial/orbital excitations, additional hybrid modes, and multi-hadron continua, are assumed to be separated from $\omega_{1,2,H}$ by energy gaps much larger than the driving Rabi frequencies $\vert{}\Omega_{1H}\vert{}$ and $\vert{}\Omega_{2H}\vert{}$, as well as the dissipation rates $\Gamma_H$ and $\gamma_{12}$. Under this timescale separation, out-of-sector states do not dynamically participate in coherent dark-state interference. Their effect is limited to static, perturbative renormalizations of the effective energies and coupling strengths within $\mathcal{H}_{\mathrm{eff}}$, in the spirit of a Feshbach–Schur projection \cite{feshbach1958}. While full quantitative validation would require explicit level spacings and transition matrix elements from lattice QCD or Born–Oppenheimer effective field theories, this three-state model offers a controlled, minimal framework that captures the dominant coherent pathways while treating higher configurations as a passive background.

	%%%%%%%%%%%%%%%%%%%%%%%%%%%
	\section{CE dipole moment and coherent response}
	\label{sec:opacity_control}
	
	The dark state introduced by equation \eqref{eq:chromoelectric_dark_state} provides a microscopic criterion for destructive interference in the reduced three-state system. However, a suppressed hybrid-state population alone is not enough to ensure an EIT-like transparency window \cite{ElectromagneticallyInducedTransparency.RevModPhys.77.633.2005,harris1997}. More precisely, transparency arises from destructive interference encoded in the off-diagonal coherence of the reduced density matrix. Therefore, the response must be derived from density-matrix dynamics and related to the coherence induced by a weak probe-like CE perturbation. In this section, we derive that response explicitly.
	
	%%%%%%%%%%%%%%%%%%%%%%%%%%%%%%%%%%%%
	\subsection{Effective CE polarization}
	\label{subsec:ce_polarization}
	
	Since the basis vectors represent fully confined configurations rather than individual colored constituents, the response is best described by the effective CE operator in equation~\eqref{eq:chromodipole_interaction}, whose expectation value is \cite{scully.1997}
	\begin{equation}
		d_{\mathrm{CE}}^{a}(t) =\operatorname{Tr}
		\left[\hat{\rho}(t)\mathbf{\hat{d}}^{\,a}_{\mathrm{eff}}\right]=\langle\Psi(t)|
		\mathbf{\hat{d}}^{\,a}_{\mathrm{eff}}|\Psi(t)\rangle.
		\label{eq:dipole_def}
	\end{equation}
	
	Using the definition \eqref{eq:ortho_operator},
	%\begin{equation}
	%d_{ij}^{a} =
	%\langle i|\hat{d}^{\,a}|j\rangle,
	%\qquad
	%i,j\in\{1,2,H\},
	%\label{eq:dipole_matrix_elements}
	%\end{equation}
	the component oscillating at the probe frequency is proportional to the coherence between the probe-coupled lower configuration $\lvert 1\rangle$ and the hybrid configuration $\lvert H\rangle$
	\begin{equation}
		d_{\mathrm{CE}}^{a}(\omega_{p}) =
		d_{1H}^{a}\rho_{H1}^{(1)} + d_{H1}^{a}\rho_{1H}^{(1)},
		\label{eq:probe_dipole_component}
	\end{equation}
	where the superscript $(1)$ denotes the term that is first order in the weak probe-like coupling $\Omega_{1H}$. For an effective ensemble density $n_{\mathrm{eff}}$, we define \cite{boyd_book}
	\begin{equation}
		\mathcal{P}_{\mathrm{CE}}^{a}(\omega_{p}) = n_{\mathrm{eff}}\,
		d_{\mathrm{CE}}^{a}(\omega_{p}),
		\label{eq:ce_polarization}
	\end{equation}
	and introduce the scalar response coefficient \cite{boyd_book}
	\begin{equation}
		\mathcal{P}_{\mathrm{CE}}(\omega_{p}) = \chi_{\mathrm{CE}}(\omega_{p})
		\, \delta\mathcal{E}_{p}(\omega_{p}).
		\label{eq:susceptibility_def}
	\end{equation}
	where $\chi_{\mathrm{CE}}$ is an effective color-response function defined by the confined-state CE operator and the probe-induced coherence in the reduced model, rather than by a literal optical medium. The analogy with optical susceptibility is operational: a decrease in $\operatorname{Im}\chi_{\mathrm{CE}}$ signifies reduced excitation of the hybrid channel within the effective confined-state dynamics. Since $\Omega_{1H}$ is proportional to the probe-field amplitude, the linear susceptibility can be expressed as
	\begin{equation}
		\chi_{\mathrm{CE}}(\omega_{p}) = \mathcal{N}_{\mathrm{CE}}
		\frac{\rho_{H1}^{(1)}}{\Omega_{1H}},
		\label{eq:chi_from_coherence}
	\end{equation}
	where $\mathcal{N}_{\mathrm{CE}}$ encompasses the effective density, the color-dipole matrix element, and the adopted normalization conventions. In direct analogy with optical EIT, the absorptive-like part of the response is therefore proportional to
	\begin{equation}
		\operatorname{Im}\chi_{\mathrm{CE}}
		\propto
		\operatorname{Im}
		\left(
		\frac{\rho_{H1}^{(1)}}{\Omega_{1H}}
		\right),
		\label{eq:absorption_from_coherence}
	\end{equation}
	and then a local minimum of $\operatorname{Im}\chi_{\mathrm{CE}}$ may be interpreted as an EIT-like transparency window, namely, a suppression of the hybrid-channel excitation induced by destructive interference in the effective three-level system.
	
	%%%%%%%%%%%%%%%%%%%%%%%%%%%%%%%%%%%%%%%%%%
	\subsection{Open-system dynamics}
	\label{subsec:open_system}
	
	Including the finite lifetime of the hybrid configuration ($\Gamma_H$) and collisional or thermal dephasing between the lower confined configurations ($\gamma_{12}$) is essential for a realistic description \cite{breuer_book}. These two rates determine the depth and contrast of the EIT-like response. Together, these dissipation parameters determine the depth, contrast, and threshold $\vert{}\Omega_{2H}\vert{}^2 \gg \Gamma_H \gamma_{12}$ needed to observe an EIT-like transparency window in CE systems. To implement this framework, we use a Markovian quantum master equation of the Gorini–Kossakowski–Sudarshan-Lindblad (GKSL) type \cite{gorini.1976,lindblad.1976}, written as
	\begin{equation}
		\dot{\hat{\rho}} = -i
		\left[\hat{H}_{\mathrm{eff}},\hat{\rho}\right]
		+ \mathcal{L}_{H}[\hat{\rho}]
		+ \mathcal{L}_{12}[\hat{\rho}],
		\label{eq:lindblad_master}
	\end{equation}
	where $\hat{H}_{\mathrm{eff}}$ is given by equation~\eqref{eq:Heff_dark}. The dissipative terms are written in GKSL form \cite{gorini.1976,lindblad.1976}
	\begin{equation}
		\mathcal{L}_{H}[\hat{\rho}] = \sum_{j=1,2}
		\left(\hat{L}_{Hj}\hat{\rho}\hat{L}_{Hj}^{\dagger}
		- \frac{1}{2} \left\{\hat{L}_{Hj}^{\dagger}\hat{L}_{Hj},
		\hat{\rho} \right\} \right),
		\label{eq:lindblad_hybrid}
	\end{equation}
	where
	\begin{equation}
		\hat{L}_{Hj} =
		\sqrt{\Gamma_{Hj}} \lvert j\rangle\langle H\rvert,
		\qquad
		\Gamma_{Hj} = \Gamma_{H1} + \Gamma_{H2},
		\label{eq:hybrid_decay_operators}
	\end{equation}
	and
	\begin{equation}
		\mathcal{L}_{12}[\hat{\rho}]= \frac{\gamma_{\varphi}}{2}\left(
		\hat{L}_{\varphi}\hat{\rho}\hat{L}_{\varphi}^{\dagger}
		-\frac{1}{2} \left\{\hat{L}_{\varphi}^{\dagger}\hat{L}_{\varphi},
		\hat{\rho}\right\}\right),
		\qquad
		\hat{L}_{\varphi}=\lvert 2\rangle\langle 2\rvert-
		\lvert 1\rangle\langle 1\rvert.
		\label{eq:lower_dephasing}
	\end{equation}
	
	%The contribution from pure dephasing comes from the action of a jump operator proportional to the population projection operator, \(L = \sqrt{2\gamma_{\varphi}} \, \vert 2\rangle\langle 2\vert\). 
	Since states $\lvert 1\rangle$ and $\lvert 2\rangle$ are orthonormal to each other, the operator $\hat{L}_{\varphi}$ is Hermitian ($\hat{L}_{\varphi}^{\dagger} = \hat{L}_{\varphi}$). Computing its square yields
	\begin{equation}
		\hat{L}_{\varphi}^{\dagger}\hat{L}_{\varphi} = (\lvert 2\rangle\langle 2\rvert - \lvert 1\rangle\langle 1\rvert)^2 = \lvert 2\rangle\langle 2\rvert + \lvert 1\rangle\langle 1\rvert = \hat{\mathbb{I}}_{12},
		\label{eq:app_L_square}
	\end{equation}
	where $\hat{\mathbb{I}}_{12}$ acts as the identity operator within the lower two-level subspace. Substituting equation~\eqref{eq:app_L_square} into equation~\eqref{eq:lower_dephasing}, the anticommutator term simplifies to $\frac{1}{2}\left\{ \hat{\mathbb{I}}_{12}, \hat{\rho} \right\} = \hat{\rho}$, giving
	\begin{equation}
		\mathcal{L}_{12}[\hat{\rho}] = \frac{\gamma_{\varphi}}{2} \left( \hat{L}_{\varphi}\hat{\rho}\hat{L}_{\varphi} - \hat{\rho} \right).
		\label{eq:app_lindblad_simplified}
	\end{equation}
	
	To isolate the rate of change of the off-diagonal coherence element $\rho_{21} \equiv \langle 2 \rvert \hat{\rho} \lvert 1 \rangle$, we project equation~\eqref{eq:app_lindblad_simplified} onto the states $\langle 2 \rvert$ and $\lvert 1 \rangle$
	\begin{equation}
		\left.\dot{\rho}_{21}\right|_{\varphi} = \langle 2 \rvert \mathcal{L}_{12}[\hat{\rho}] \lvert 1 \rangle = \frac{\gamma_{\varphi}}{2} \left( \langle 2 \rvert \hat{L}_{\varphi}\hat{\rho}\hat{L}_{\varphi} \lvert 1 \rangle - \langle 2 \rvert \hat{\rho} \lvert 1 \rangle \right).
		\label{eq:app_rho21_projection}
	\end{equation}
	
	Using the action of the jump operator on the basis states, $\hat{L}_{\varphi}\lvert 1\rangle = -\lvert 1\rangle$ and $\langle 2\rvert\hat{L}_{\varphi} = \langle 2\rvert$, the first term in the brackets evaluates to
	\begin{equation}
		\langle 2 \rvert \hat{L}_{\varphi}\hat{\rho}\hat{L}_{\varphi} \lvert 1 \rangle = (\langle 2 \rvert \hat{L}_{\varphi}) \hat{\rho} (\hat{L}_{\varphi} \lvert 1 \rangle) = \langle 2 \rvert \hat{\rho} (-)\lvert 1 \rangle = -\rho_{21}.
	\end{equation}
	
	Substituting this result back into equation~\eqref{eq:app_rho21_projection}, we obtain 
	\begin{equation}
		\left.\dot{\rho}_{21}\right|_{\varphi} = \frac{\gamma_{\varphi}}{2} \left( -\rho_{21} - \rho_{21} \right) = -\gamma_{\varphi}\rho_{21}.
	\end{equation}
	
	%With the convention adopted in equation~\eqref{eq:lower_dephasing}, the pure-dephasing contribution satisfies (see Appendix C)
	%\begin{equation}
	%\left.\dot{\rho}_{21}\right|_{\varphi} =
	%-\gamma_{\varphi}\rho_{21},
	%\label{eq:pure_dephasing_rate}
	%\end{equation}
	Using the above results, the total lower-state decoherence rate is written as
	\begin{equation}
		\gamma_{12} = \gamma_{\varphi}
		+ \frac{\Gamma_{1\rightarrow2}+\Gamma_{2\rightarrow1}}{2}.
		\label{eq:gamma12_definition}
	\end{equation}
	
	In the numerical results below, no incoherent transfer between the lower configurations is assumed, so that $\gamma_{12}=\gamma_{\varphi}$. Moreover, we work in the standard weak-probe regime (to prevent disturbance of the quantum medium or the introduction of nonlinear effects), where
	\begin{equation}
		\left|\Omega_{1H}\right|\ll\left|\Omega_{2H}
		\right|,
		\label{eq:weak_probe_condition}
	\end{equation}
	and assume that, to zeroth order in $\Omega_{1H}$, there is an almost complete ground state and the absence of residual population in the state $\rvert 2\rangle$ or thermal excitation in the state $\rvert H\rangle$. These conditions are respectively written as 
	\begin{equation}
		\rho_{11}^{(0)} \simeq 1,
		\qquad
		\rho_{22}^{(0)} \simeq \rho_{HH}^{(0)} \simeq 0.
		\label{eq:zeroth_order_populations}
	\end{equation}

	In the standard EIT protocol \cite{breuer_book}, the probe-like perturbation is much weaker than the coupling-like field ($|\Omega_{1H}| \ll |\Omega_{2H}|$). This means that we can perform a perturbative expansion of the density matrix in powers of the probe amplitude $\Omega_{1H}$
	\begin{equation}
		\hat{\rho} = \hat{\rho}^{(0)} + \hat{\rho}^{(1)} + \mathcal{O}(\Omega_{1H}^2),
	\end{equation}
	%%%%%%%%%%%%%%%%%%%%%%%%%%%
	%\subsubsection{Zeroth-Order Unperturbed State}
	and in the absence of the probe field ($\Omega_{1H} = 0$), the system relaxes to the ground configuration $\lvert 1\rangle$. Consequently, the zeroth-order steady-state density matrix is given by
	\begin{equation}
		\hat{\rho}^{(0)} = \begin{pmatrix} 1 & 0 & 0 \\ 0 & 0 & 0 \\ 0 & 0 & 0 \end{pmatrix},
	\end{equation}
	which implies the populations and coherences at order $\mathcal{O}(1)$ satisfy $\rho_{11}^{(0)} = 1$, $\rho_{22}^{(0)} = 0$, $\rho_{HH}^{(0)} = 0$, and $\rho_{ij}^{(0)} = 0$ for $i \neq j$.
	
	%%%%%%%%%%%%%%%%%%%%%%%%%\eqref{eq:bloch_H1} and \eqref{eq:bloch_21}
	%\subsubsection{First-Order Linearized Coherences}
	%Substituting the expansion into equations~\eqref{eq:bloch_H1}%{eq:app_rhoH1_exact} 
	%and \eqref{eq:bloch_21}%{eq:app_rho21_exact} 
	%and collecting terms strictly linear in $\Omega_{1H}$ yields:
	
	Considering the case for the optical coherence $\rho_{H1}^{(1)}$, one obtains
	\begin{equation}
		\dot{\rho}_{H1}^{(1)} = -\left(\frac{\Gamma_H}{2} - i\Delta_H\right)\rho_{H1}^{(1)} + \frac{i\Omega_{1H}}{2} \left(\rho_{11}^{(0)} - \rho_{HH}^{(0)}\right) + \frac{i\Omega_{2H}}{2} \rho_{21}^{(1)},
	\end{equation}
	and using $\rho_{11}^{(0)} - \rho_{HH}^{(0)} = 1 - 0 = 1$, this simplifies to
	\begin{equation}
		\dot{\rho}_{H1}^{(1)} = -\left(\frac{\Gamma_H}{2} - i\Delta_H\right)\rho_{H1}^{(1)} + \frac{i\Omega_{1H}}{2} + \frac{i\Omega_{2H}}{2} \rho_{21}^{(1)}.
		\label{eq:bloch_H1}
	\end{equation}
	%which proves equation~\eqref{eq:bloch_H1}.
	
	The first term on the right-hand side of equation \eqref{eq:bloch_H1} represents damping and detuning; the second corresponds to direct excitation, and the third accounts for coherence transfer, or coupling-induced coherence transfer. On the other hand, for the two-photon coherence $\rho_{21}^{(1)}$, one has
	\begin{equation}
		\dot{\rho}_{21}^{(1)} = -\left(\gamma_{12} - i\delta\right)\rho_{21}^{(1)} + \frac{i\Omega_{2H}^*}{2} \rho_{H1}^{(1)} - \frac{i\Omega_{1H}}{2} \rho_{2H}^{(0)},
	\end{equation}
	and since $\rho_{2H}^{(0)} = 0$ at zeroth order, the last term vanishes completely, leaving
	\begin{equation}
		\dot{\rho}_{21}^{(1)} = -\gamma_{12}\rho_{21}^{(1)} + i\delta\rho_{21}^{(1)} + \frac{i\Omega_{2H}^*}{2} \rho_{H1}^{(1)}.
		\label{eq:bloch_21}
	\end{equation}
	%which establishes Eq.~\eqref{eq:bloch_21}.
	%The probe-induced coherences $\rho_{H1}^{(1)}$ and $\rho_{21}^{(1)}$ are connected throughout the pair of equations (see Appendix D)
	%\begin{align}
	%\dot{\rho}_{H1}^{(1)} &= - \left(\frac{\Gamma_{H}}{2}- i\Delta_{H} \right) \rho_{H1}^{(1)} + \frac{i\Omega_{1H}}{2} + \frac{i\Omega_{2H}}{2} \rho_{21}^{(1)},
	%\label{eq:bloch_H1}
	%\\
	%\dot{\rho}_{21}^{(1)} &= -
	%\gamma_{12}\rho_{21}^{(1)} + i\delta\rho_{21}^{(1)} + \frac{i\Omega_{2H}^{\ast}}{2} \rho_{H1}^{(1)}.
	%\label{eq:bloch_21}
	%\end{align}
	where the first term represents decay of the two-photon coherence due to environmental dephasing, the second accounts for phase oscillation driven by the two-photon detuning, and the third describes coherence transfer from the excited state via the coupling field. Taking \eqref{eq:weak_probe_condition} into account, second-order and higher terms in the probe coupling have been omitted in equations \eqref{eq:bloch_H1}-\eqref{eq:bloch_21}. In particular, $\rho_{H2}^{(1)}$ does not contribute to the linear probe response under the population assumption in relation~\eqref{eq:zeroth_order_populations}.
	
	%%%%%%%%%%%%%%%%%%%%%%%%%%%%%%%%%%%%%%%%%%%%%
	\subsection{Steady-state coherence and susceptibility}
	\label{subsec:steady_state_susceptibility}
	
	The stationary regime imposes the condition
	\begin{equation}
		\dot{\rho}_{H1}^{(1)} = \dot{\rho}_{21}^{(1)} = 0,
		\label{eq:steady_state_condition}
	\end{equation}
	while equation~\eqref{eq:bloch_21} gives
	\begin{equation}
		\rho_{21}^{(1)} = \frac{
			i\Omega_{2H}^{\ast}}{2\left(\gamma_{12}-i\delta\right)}
		\rho_{H1}^{(1)}.
		\label{eq:rho21_solution}
	\end{equation}
	
	Now, replacing equation \eqref{eq:rho21_solution} into equation~\eqref{eq:bloch_H1} yields
	\begin{equation}
		\left[ \left( \frac{\Gamma_{H}}{2} -
		i\Delta_{H} \right) + \frac{|\Omega_{2H}|^{2}}{4\left(
			\gamma_{12}-i\delta\right)}\right]
		\rho_{H1}^{(1)} = \frac{i\Omega_{1H}}{2}.
		\label{eq:rhoH1_intermediate}
	\end{equation}
	and, therefore, one obtains
	\begin{equation}
		%\boxed{
			\rho_{H1}^{(1)} = \frac{i\Omega_{1H} \left(
				\gamma_{12}-i\delta\right)}{2\left[\left(\frac{\Gamma_{H}}{2}
				- i\Delta_{H} \right)\left(\gamma_{12}-i\delta\right)
				+ \frac{|\Omega_{2H}|^{2}}{4} \right]}
			%}
		\label{eq:rhoH1_solution}
	\end{equation}
	and the effective CE susceptibility becomes
	%\begin{equation}
	%\boxed{
		%\chi_{\mathrm{CE}}(\Delta_{H},\delta) = \frac{i\mathcal{N}_{\mathrm{CE}}
			%\left(\gamma_{12}-i\delta\right)}{2\left[\left(\frac{\Gamma_{H}}{2}
			%- i\Delta_{H} \right)\left(\gamma_{12}-i\delta\right)
			%+ \frac{|\Omega_{2H}|^{2}}{4}\right]}
		%}.
	%\label{eq:chi_CE_derived}
	%\end{equation}
	\begin{equation}
		%\boxed{
			\operatorname{Im} \chi_{\mathrm{CE}}
			= \frac{\mathcal{N}_{\mathrm{CE}}\left[\frac{\Gamma_{H}}{2}
				\left(\delta^{2}+\gamma_{12}^{2}\right)
				+
				\frac{\gamma_{12}|\Omega_{2H}|^{2}}{4}\right]}{2
				\left(C^{2}+D^{2}\right)},
			%}.
		\label{eq:im_chi_CE}
	\end{equation}
	where one defines
	\begin{align}
		C = \frac{\Gamma_{H}\gamma_{12}}{2} - \Delta_{H}\delta
		+ \frac{|\Omega_{2H}|^{2}}{4},
		\qquad
		D = \frac{\Gamma_{H}\delta}{2} + \gamma_{12}\Delta_{H}.
	\end{align}
	%The absorptive-like part of the response is then
	
	Equation~\eqref {eq:im_chi_CE} shows the EIT-like line shape from the reduced CE model, not a phenomenological product of a resonant envelope and a suppression factor. At exactly two-photon resonance, $\delta=0$, equation~%\eqref{eq:chi_CE_derived}
	\eqref{eq:im_chi_CE} reduces to
	\begin{equation}
		\chi_{\mathrm{CE}}(\Delta_{H},0) =
		\frac{i\mathcal{N}_{\mathrm{CE}}\gamma_{12}}{2
			\left[\gamma_{12}\left(\frac{\Gamma_{H}}{2}-
			i\Delta_{H}\right) + \frac{|\Omega_{2H}|^{2}}{4}\right]},
		\label{eq:chi_two_photon}
	\end{equation}
	while at simultaneous one- and two-photon resonance, one has
	\begin{equation}
		\Delta_{H} = \delta=0,
		\label{eq:double_resonance}
	\end{equation}
	resulting in the residual absorptive response
	\begin{equation}
		\operatorname{Im}
		\chi_{\mathrm{CE}}(0,0) = \frac{\mathcal{N}_{\mathrm{CE}}\gamma_{12}}{
			2\left[\frac{\Gamma_{H}\gamma_{12}}{2}+\frac{|\Omega_{2H}|^{2}}{4}
			\right]}.
		\label{eq:residual_absorption}
	\end{equation}
	
	Consequently, the following result can be easily obtained
	\begin{equation}
		\lim_{\gamma_{12}\rightarrow 0} \operatorname{Im}
		\chi_{\mathrm{CE}}(0,0) = 0,
		\label{eq:derived_transparency_limit}
	\end{equation}
	showing that the transparency window is a consequence of the vanishing lower-state decoherence and destructive interference at the two-photon resonance ($\delta = 0$). For finite $\gamma_{12}$, the suppression is incomplete, and a residual absorption remains. As a control, when the coupling-like field is removed,
	\begin{equation}
		\Omega_{2H} = 0,
		\label{eq:no_control_field}
	\end{equation}
	implying that equation~%\eqref{eq:chi_CE_derived}
	\eqref{eq:im_chi_CE} reduces to the two-level response
	\begin{equation}
		\chi_{\mathrm{CE}}^{(2)} =
		\frac{i\mathcal{N}_{\mathrm{CE}}}{2
			\left(\frac{\Gamma_{H}}{2}-i\Delta_{H}\right)},
		\label{eq:two_level_susceptibility}
	\end{equation}
	which has a maximum absorptive response at $\Delta_{H}=0$ and no transparency dip. This comparison isolates the role of the second coherent pathway. Meanwhile, the hybrid-state population
	\begin{equation}
		P_{H} = \rho_{HH}
		\label{eq:hybrid_population_suppression}
	\end{equation}
	remains a useful indicator of coherent population trapping, but the transparency criterion is now derived from equation~\eqref{eq:im_chi_CE}. In the remainder of the manuscript, the absorptive-like response is therefore
	\begin{equation}
		\mathcal{A}_{\mathrm{CE}} \left( \Delta_{H},\delta \right)
		= \operatorname{Im} \chi_{\mathrm{CE}} \left(\Delta_{H},\delta
		\right),
		\label{eq:derived_absorption_response}
	\end{equation}
	instead of an independently postulated model for $P_H$. 
	
	It may be useful to note that, within medium-induced energy-loss formalisms, the opacity of the quark-gluon medium is often characterized by the imaginary part of the color response functions. The appearance of a vanishing imaginary susceptibility $\mathrm{Im}\,\chi_{CE} \to 0$ at two-photon resonance could indicate a highly anisotropic, frequency-selective color opacity. If collective soft-gluon modes in heavy-ion collisions act as the effective coupling field $\Omega_{2H}$, the color medium may become dynamically transparent to probe perturbations at specific frequencies $\omega_p$. Such a modification of the spectral density could, in turn, alter the energy-loss rates of heavy-flavor probes traversing regions with coherent CE backgrounds.
	
	%%%%%%%%%%%%%%%%%%%%%%%%%%%%%%%%%%%%%%%%%%%%%%%%%%%%
	\section{Control cases for the EIT-like CE response}
	\label{sec:control_cases}
	
	%%%%%%%%%%%%%%%%%%%%%%%%%%%%%%%%%%%%%%%
	%\subsection{Calibration of Phenomenological Non-Abelian Parameters from Charmonium Lattice Data}
	%\label{subsec:parameter_calibration}
	
	Lattice simulations of pure $SU(3)$ gauge theory with static sources \cite{dudek2008} indicate that the $\Sigma_{u}^{-}$ hybrid state lies about $1.0 - 1.2$ GeV above the conventional $\Sigma_{g}^{+}$ ground state at interquark separations relevant to flux-tube formation ($r \sim 0.5 - 1.0$ fm). This splitting provides the natural lattice-calibration scale for the present framework: the non-Abelian correction $\Delta\omega_{H}^{\mathrm{NA}}(r,s)$ enters as a diagonal self-energy shift on $|H\rangle$, so it must remain parametrically smaller than the observed hybrid excitation gap in order not to distort the ordering and isolation of the low-lying Born-Oppenheimer sectors. Accordingly, we require $\Delta\omega_{H}^{\mathrm{NA}} \ll 1$ GeV, which is a conservative numerical consistency condition rather than an arbitrary bound; if the shift were comparable to the physical level splitting, it would significantly renormalize the hybrid energy, drive the state toward resonance with higher-lying gluoexcitations or continua, and thereby undermine the assumptions underlying the Feshbach-Schur truncated Hilbert space \cite{feshbach1958}.
	
	Consequently, in the absence of a first-principles Yang-Mills derivation for the conversion factor $\kappa$, we adopted illustrative numerical benchmarks to ensure that $\Delta\omega_{H}^{\mathrm{NA}}$ remains well within the sub-GeV range. For a representative flux-tube radius $s=0.01$ fm and standard strong coupling $g$, $\kappa$ was chosen so that the correction, of order $10^{-1}$ GeV, produces a visible spectral shift while preserving the hierarchy required for dark-state formation: $|\Omega_{2H}| \sim \Gamma_{H} \gg \Delta\omega_{H}^{\mathrm{NA}}$. 
	
	To make the analogy with EIT more precise, we introduce control scenarios to provide a reference for quantitative comparison with the CE model. These cases are not meant to realize EIT in a realistic QGP system, but to identify which features of the effective confined color model are analogous to optical EIT and which arise from the non-Abelian structure of the CE interaction.
	
	We analyze three reference configurations: the conventional optical $\Lambda$-type EIT system, the effective CE three-level system without the non-Abelian hybrid-energy correction, and the full CE model with the non-Abelian correction. Comparing these cases clarifies the roles of flux-tube geometry and the non-Abelian gluonic sector in the emergent EIT-like response.
	
	The three cases are described by the same $\Lambda$-type Hamiltonian given by equation~\eqref{eq:Heff_dark}. The probe- and coupling-like Rabi frequencies are
	\begin{equation}
		\Omega_{1H}(r,s) = -g\,\mu_{1H}\,\mathcal{F}(r,s),
		\qquad
		\Omega_{2H}(r,s) = -g\,\mu_{2H}\,\mathcal{F}(r,s),
		\label{eq:control_rabi_frequencies}
	\end{equation}
	where $\mu_{1H}$ and $\mu_{2H}$ are the effective color-dipole matrix elements defined in equation~\eqref{eq:effective_overlap}. The rotating-frame Hamiltonian is therefore given by equation \eqref{eq:Heff_dark}
	%\begin{equation}
	%\hat{H}_{\mathrm{CE}} = \frac{1}{2}
	%\begin{pmatrix}
	%0 & 0 & \Omega_{1H} \\
	%0 & 2\delta & \Omega_{2H} \\
	%\Omega_{1H}^{\ast} & \Omega_{2H}^{\ast} & 2\Delta_{H}
	%\end{pmatrix},
	%\label{eq:H_CE_control}
	%\end{equation}
	in the ordered basis \eqref{eq:ord_basis}. 
	%\begin{equation}
	%\left\{
	%\lvert 1\rangle,
	%\lvert 2\rangle,
	%\lvert H\rangle
	%\right\}.
	%\end{equation}
	The corresponding detunings are
	\begin{align}
		\Delta_{1} &= \omega_{p} - \left(\omega_{H}-\omega_{1}\right),
		\\
		\Delta_{2} &= \omega_{c} - \left(\omega_{H}-\omega_{2}\right),
		\\
		\delta &= \Delta_{1} - \Delta_{2},
		\qquad
		\Delta_{H}\equiv\Delta_{1}.
		\label{eq:control_detunings}
	\end{align}
	
	The optical benchmark is obtained by interpreting the quantities in equation~\eqref{eq:Heff_dark} as ordinary optical Rabi frequencies, detunings, and decay rates. The CE reference model uses the confined-state parameters but sets
	\begin{equation}
		\Delta\omega_{H}^{\mathrm{NA}}(s) = 0,
		\label{eq:no_nonabelian_shift}
	\end{equation}
	while the full CE model instead uses
	\begin{equation}
		\omega_{H}^{\mathrm{self}}(r,s) =
		\omega_{H}^{(0)}(r,s) + \Delta\omega_{H}^{\mathrm{NA}}(s),
		\qquad
		\Delta\omega_{H}^{\mathrm{NA}}(s) = \kappa\mathcal{E}_{\perp}(s),
		\label{eq:full_nonabelian_shift}
	\end{equation}
	as defined in equation~\eqref{eq:nonabelian_shift}. Thus, the non-Abelian contribution represents the control comparison as a diagonal shift in the hybrid-state energy, not as an additive term in the transition amplitudes. Then, the non-Abelian self-interaction renormalizes the gluon-excited hybrid state’s self-energy, appearing as a diagonal energy shift $\Delta\omega_H^{\mathrm{NA}}(r,s)$ in the effective Hamiltonian. Since this correction does not change the off-diagonal transition matrix elements $\Omega_{1H}$ and $\Omega_{2H}$, the non-Abelian medium effects do not undermine the quantum interference behind dark-state formation. Alternatively, they induce a purely dispersive renormalization of the one-photon detuning $\Delta_H \to \Delta_H + \Delta\omega_H^{\mathrm{NA}}$, shifting the resonance peak while preserving the EIT-like transparency condition at $\delta = 0$. For simplicity, we define the single-field (or single-quantum) detuning $\Delta_H = \omega_p-\omega_{H1}$, conventionally referred to in EIT literature as the one-photon detuning \cite{fleishhauer2000}.
	
	The comparison between the control cases is now performed using the derived absorptive-like response given by equation~\eqref{eq:derived_absorption_response}
	%\begin{equation}
	%\mathcal{A}_{\mathrm{CE}}\left(\Delta_{H},\delta\right)= \operatorname{Im} \chi_{\mathrm{CE}} \left(\Delta_{H},\delta \right),
	%\label{eq:absorption_control_cases}
	%\end{equation}
	with $\mathrm{Im}\chi_{\mathrm{CE}}$ given by equation~\eqref{eq:im_chi_CE}. %{eq:chi_CE_derived}.  
	The optical benchmark has the same functional structure, with the CE parameters replaced by their optical counterparts.
	
	For a one-dimensional detuning scan, the coupling-like field is kept on resonance ($\Delta_2=0$, while the probe frequency is varied. Since $\delta=\Delta_1-\Delta_2$ and $\Delta_H\equiv\Delta_1$, it follows that
	\begin{equation}
		\Delta_H = \delta\equiv \Delta,
		\label{eq:one_dimensional_scan}
	\end{equation}
	and equation~\eqref{eq:chi_two_photon} %{eq:im_chi_CE}%{eq:chi_CE_derived} 
	gives
	\begin{equation}
		\chi_{\mathrm{CE}}(\Delta)=
		\frac{i\mathcal{N}_{\mathrm{CE}}\left(
			\gamma_{12}-i\Delta\right)}{2
			\left[\left(\frac{\Gamma_{H}}{2}-i\Delta\right)
			\left(\gamma_{12}-i\Delta\right)
			+\frac{|\Omega_{2H}|^{2}}{4}\right]}.
		\label{eq:chi_one_dimensional_scan}
	\end{equation}
	
	To decouple the characteristic spectral profile, contrast, and depth of the EIT-like transparency window from the absolute magnitude of the CE coupling, we use equation \eqref{eq:derived_absorption_response} to introduce the normalized absorption spectrum $\widetilde{\mathcal{A}}_{\mathrm{CE}}(\Delta)$, defined as
	\begin{equation}
		\widetilde{\mathcal{A}}_{\mathrm{CE}}(\Delta) = \frac{\operatorname{Im}\chi_{\mathrm{CE}}(\Delta)}{\max_{\Delta}\operatorname{Im}\chi_{\mathrm{CE}}(\Delta)} \in [0,1].
		\label{eq:normalized_absorption}
	\end{equation}
	where $\widetilde{\mathcal{A}}_{\mathrm{CE}} = 1$ corresponds to maximal opacity (i.e., peak absorption or dissipation), whereas $\widetilde{\mathcal{A}}_{\mathrm{CE}} = 0$ indicates ideal transparency arising from the formation of a dark state. The control model without a coherent coupling pathway is obtained by setting
	\begin{equation}
		\Omega_{2H}=0,
		\label{eq:control_without_coupling}
	\end{equation}
	which yields the two-level Lorentzian response in equation~\eqref{eq:two_level_susceptibility}. The EIT-like model retains $\Omega_{2H}\neq0$, producing the central transparency dip derived above.
	
	%The non-Abelian correction affects the response through the shifted hybrid energy,
	%\begin{equation}
	%\omega_{H}^{\mathrm{self}}(r,s)= \omega_{H}^{(0)}(r,s)
	%+\Delta\omega_{H}^{\mathrm{NA}}(r,s),
	%\label{eq:hybrid_energy_control}
	%\end{equation}
	%and hence through the shifted one-photon detuning
	%\begin{equation}
	%\Delta_{H}^{\mathrm{self}} = \Delta_{H}^{0}
	%- \Delta\omega_{H}^{\mathrm{NA}}(s).
	%\label{eq:detuning_shift_control}
	%\end{equation}
	
	To quantify how the non-Abelian gluon self-interaction alters the absorption profile, we introduce the differential response $\Delta\mathcal{A}_{\mathrm{CE}}(\Delta)$, defined as the difference between the full dressed response and the reference unshifted profile
	\begin{equation}\Delta\mathcal{A}{\mathrm{CE}}(\Delta) =
		\widetilde{\mathcal{A}}{\mathrm{CE}}^{,\mathrm{self}}(\Delta)-\widetilde{\mathcal{A}}_{\mathrm{CE}}^{,0}(\Delta).
		\label{eq:delta_absorption_self}
	\end{equation}
	
	In this formulation, the transparency minimum shifts due to the non-Abelian correction to the hybrid energy in equation~\eqref{eq:hybrid_energy_correction}, while the minimum depth is determined mainly by $\gamma_{12}$ and $|\Omega_{2H}|$. To isolate the spectral region dominated by dark-state formation, the transparency-window width $\Gamma_{\mathrm{tr}}$ is defined as the frequency interval where the normalized absorption falls below a threshold $\eta$.
	\begin{equation}
		\widetilde{\mathcal{A}}_{\mathrm{CE}}(\Delta) < \eta,
		\qquad
		(0 < \eta < 1).
		\label{eq:Gamma_threshold}
	\end{equation}
	
	For example, $\eta=0.1$ identifies the interval where the absorptive-like response stays below $10\%$ of its maximum. The net broadening or narrowing of the transparency window caused by medium self-interaction is quantified by the differential bandwidth, written as
	\begin{equation}
		\Delta\Gamma_{\mathrm{tr}} = \Gamma_{\mathrm{tr}}^{\mathrm{self}} - \Gamma_{\mathrm{tr}}^{0}.
		\label{eq:delta_Gamma}
	\end{equation}
	\begin{table}[t!]
		\centering
		\caption{Control cases separating the interference mechanism from the non-Abelian hybrid-energy correction. (1) Shows the resonant response without a second coherent pathway. (2) Provides the standard EIT line shape; (3) Isolates dark-state interference in the confined-state model; (4) Tests the spectral displacement generated by the hybrid-energy correction.}
		\label{tab:control_cases}
		\begin{tabular}{c|cc}%{p{0.23\linewidth}|p{0.30\linewidth}p{0.34\linewidth}}
			\hline
			\textbf{Case} & \textbf{Condition} & \textbf{Purpose} \\
			\hline
			Two-level reference & $\Omega_{2H}=0$ & (1) \\
			\hline
			Optical $\Lambda$ benchmark & $\Omega_p\neq0$, $\Omega_c\neq0$ & (2) \\
			\hline
			CE $\Lambda$ model without non-Abelian shift & $\Omega_{2H}\neq0$, $\Delta\omega_H^{\mathrm{NA}}=0$ & (3) \\
			\hline
			CE $\Lambda$ model with non-Abelian shift & $\Omega_{2H}\neq0$, $\Delta\omega_H^{\mathrm{NA}}\neq0$ & (4) \\
			\hline
		\end{tabular}
	\end{table}
	
	The four control cases may therefore be summarized in Table \ref{tab:control_cases}. If the CE system reproduces the normalized suppression pattern of the optical benchmark, the model supports an EIT-like coherent response. If including $\mathcal{E}_{\perp}(s)$ shifts the minimum of $\widetilde{\mathcal{A}}_{\mathrm{CE}}(\Delta)$ or changes $\Gamma_{\mathrm{tr}}$, then the non-Abelian correction has a discernible effect within the effective model. Thus, comparing the two CE control cases distinguishes a generic three-level interference effect from a specifically non-Abelian contribution linked to gluon self-coupling.
	
	In natural units ($\hbar = c = 1$), the transition rates and Rabi frequencies satisfy
	\begin{equation}
		[\Omega_{kH}] = [\Gamma_H] = [\gamma_{12}] = \mathrm{GeV},
	\end{equation}
	and since the coupling constant $g$ is dimensionless, equation~(\ref{eq:rabi_geometric_profile}) imposes the dimensional requirement
	\begin{equation}
		[\mu_{kH}] [\mathcal{F}] = \mathrm{GeV}.
		\label{eq:dimensional_condition}
	\end{equation}
	
	To avoid assigning an absolute QCD scale to $\mathcal{F}(r,s)$ before the transition matrix elements are independently determined, the spatial dependence of the coupling is introduced via the normalized flux-tube geometry
	\begin{equation}
		\Omega_{2H}(r,s) = \Omega_{2H}^{(0)} \frac{\mathcal{F}(r,s)}{\mathcal{F}(r_0,s_0)},
		\label{eq:normalized_geometric_coupling}
	\end{equation}
	which explicitly ensures that $\Omega_{2H}(r,s)$ preserves the energy/frequency dimensions across the spatial profile.

	%%%%%%%%%%%%%%%%%%%%%%%%%%%%%%%%%%%%%%
	%\subsection{Dimensionless benchmark parameters}
	%\label{subsec:benchmark_parameters}
	
	\begin{table}[t!]
		\centering
		\caption{Benchmark parameters used to evaluate the susceptibility-like response, normalized in units of $\Gamma_H$.}
		\label{tab:model_parameters}
		\begin{tabular}{c|cc}%{p{0.20\linewidth}p{0.48\linewidth}p{0.20\linewidth}}
			\hline
			\textbf{Parameter} & \textbf{Meaning} & \textbf{Value / Unit} \\
			\hline
			$\Gamma_H$ & Hybrid-state decay rate & $1.0 \; \Gamma_H$ \\
			$\gamma_{12}$ & Lower-state decoherence rate & $0.02 \; \Gamma_H$ \\
			$\Omega_{1H}$ & Weak probe-like Rabi frequency ($g \mu_{1H} \mathcal{F}$) & $0.05 \; \Gamma_H$ \\
			$\Omega_{2H}$ & Strong coupling-like Rabi frequency ($g \mu_{2H} \mathcal{F}$) & $0.60 \; \Gamma_H$ \\
			$\mathcal{N}_{\mathrm{CE}}$ & Overall response normalization factor & $1.0$ (Dimensionless) \\
			$\Delta\omega_H^{\mathrm{NA}}$ & Illustrative non-Abelian energy shift & $0.25 \; \Gamma_H$ \\
			\hline
		\end{tabular}
	\end{table}

	%%%%%%%%%%%%%%%%%%%%%%%%%%%%%%%%%%%%%%%%%%%%%
	%\subsection{Dimensional consistency}
	%\label{subsec:dimensional_consistency}
	
	%Under this convention, $\mu_{kH}$ is dimensionless. Of course, other normalizations may redistribute dimensions between $\mu_{kH}$ and $\mathcal{F}$, but their product must always satisfy equation~\eqref{eq:dimensional_condition}. 
	
	%%%%%%%%%%%%%%%%%%%%%%%%%%%%%%%%%%%%%%%%%%%%%
	\section{Theoretical predictions}
	\label{sec:theo_pred}
	
	We stress that all numerical results obtained here are theoretical predictions based on the derived line shape and are not fitted QCD predictions. All frequencies and decay rates are in units of $\Gamma_H$, with $\Gamma_H=1$ GeV. The weak-probe condition $|\Omega_{1H}|\ll|\Omega_{2H}|$ is explicitly satisfied. Table~\ref{tab:model_parameters} lists the parameter values used in this work.
	\begin{figure}[ht]
		\centering
		\includegraphics[width=0.7\linewidth]{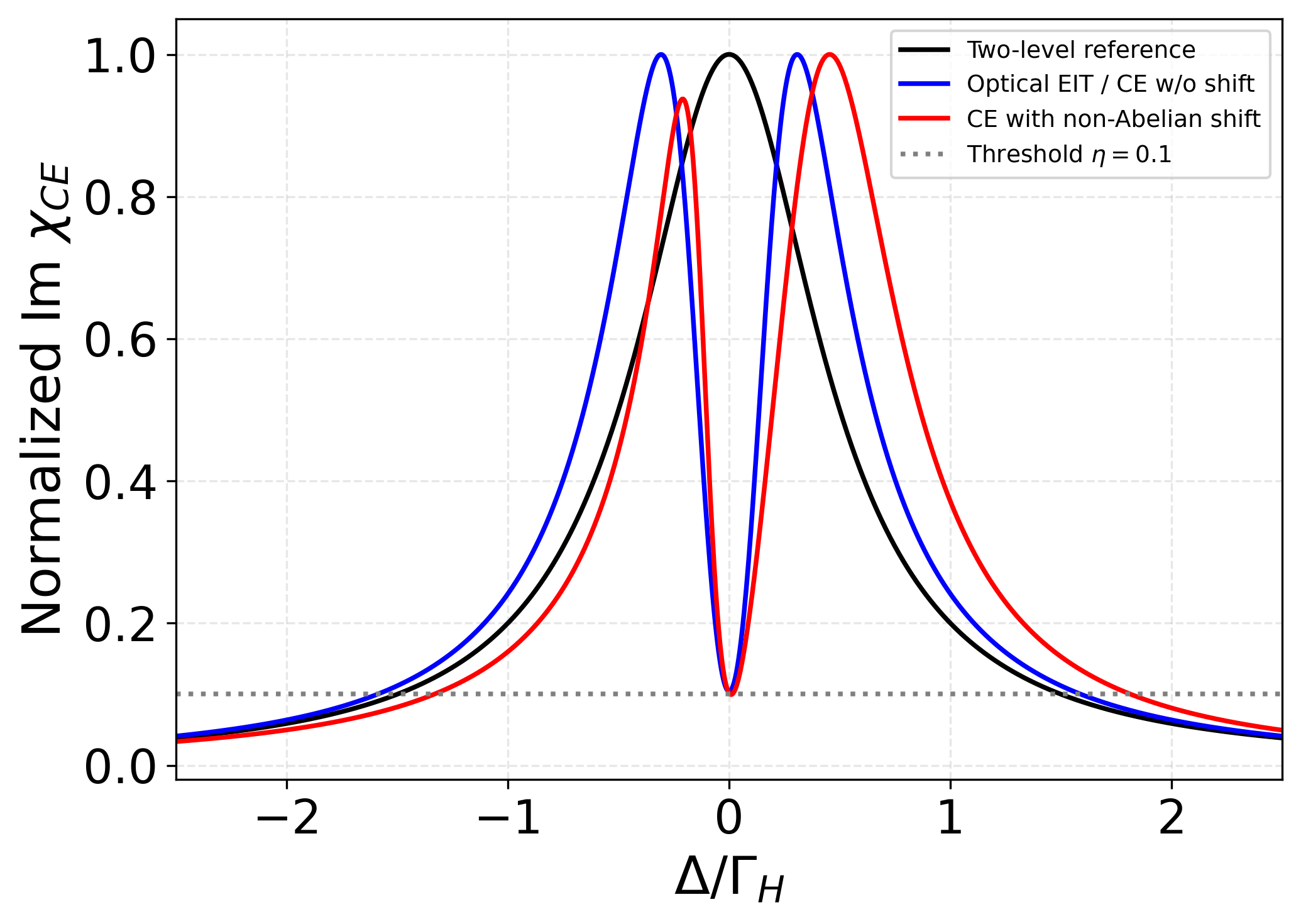}
		\caption{Normalized EIT-like response $\tilde{\mathcal{A}}_{\text{CE}}(\Delta) = \operatorname{Im}\chi_{\text{CE}}(\Delta) / \max\operatorname{Im}\chi_{\text{CE}}$ versus the dimensionless detuning $\Delta/\Gamma_H$ for the control cases in Table II. The two-level reference curve ($\Omega_{2H} = 0$) shows the standard Lorentzian profile with maximum absorption at resonance ($\Delta = 0$). The optical $\Lambda$-type benchmark and the CE model without non-Abelian shift ($\Delta\omega_H^{\text{NA}} = 0$) show a transparency window at $\Delta = 0$ due to dark-state formation. The full CE model with non-Abelian coupling ($\Delta\omega_H^{\text{NA}} \neq 0$) shows stronger suppression and reshaping of the transparency dip. The horizontal dotted line marks the operational threshold $\eta = 0.1$ used to evaluate the transparency-window width $\Gamma_{\text{tr}}$. Dimensionless parameters: $\gamma_{12}/\Gamma_H = 0.02$, $\Omega_{1H}/\Gamma_H = 0.05$, and $\Omega_{2H}/\Gamma_H = 0.60$.}
		\label{fig:normalized_response}
	\end{figure}
	
	%The normalized CE response $\tilde{\mathcal{A}}_{\text{CE}}(\Delta)$ in Figure~\ref{fig:normalized_response} for the control benchmarks in Table~\ref{tab:control_cases}. In the two-level reference limit ($\Omega_{2H}=0$), the response shows a standard Lorentzian absorption profile centered at resonance ($\Delta=0$). By contrast, the optical $\Lambda$-type benchmark shows sharp suppression of normalized absorption at $\Delta=0$, indicating two-photon resonance and dark-state formation.  The two CE model configurations reproduce this characteristic: the response remains resonant away from the destructive-interference region but shows pronounced local suppression centered at $\Delta=0$. This agreement confirms that the effective CE formulation supports an EIT-like transparency window, as defined by the operational threshold $\eta=0.1$ in equation~\eqref{eq:Gamma_threshold}. 
	
	The comparison shown in Figure~\ref{fig:normalized_response} shows how the non-Abelian correction alters the line-shape geometry: without the self-interaction shift ($\Delta\omega_H^{\text{NA}}=0$), the baseline optical dip is recovered, whereas the full CE model ($\Delta\omega_H^{\text{NA}}\neq 0$) produces a deeper, spectrally reshaped transparency minimum. %In this parameterization, the non-Abelian hybrid-energy contribution strengthens dark-state suppression near resonance, narrowing and deepening the central dip while shifting absorptive weight to the neighboring spectral shoulders.
	\begin{figure}[ht]
		\centering
		\includegraphics[width=0.7\linewidth]{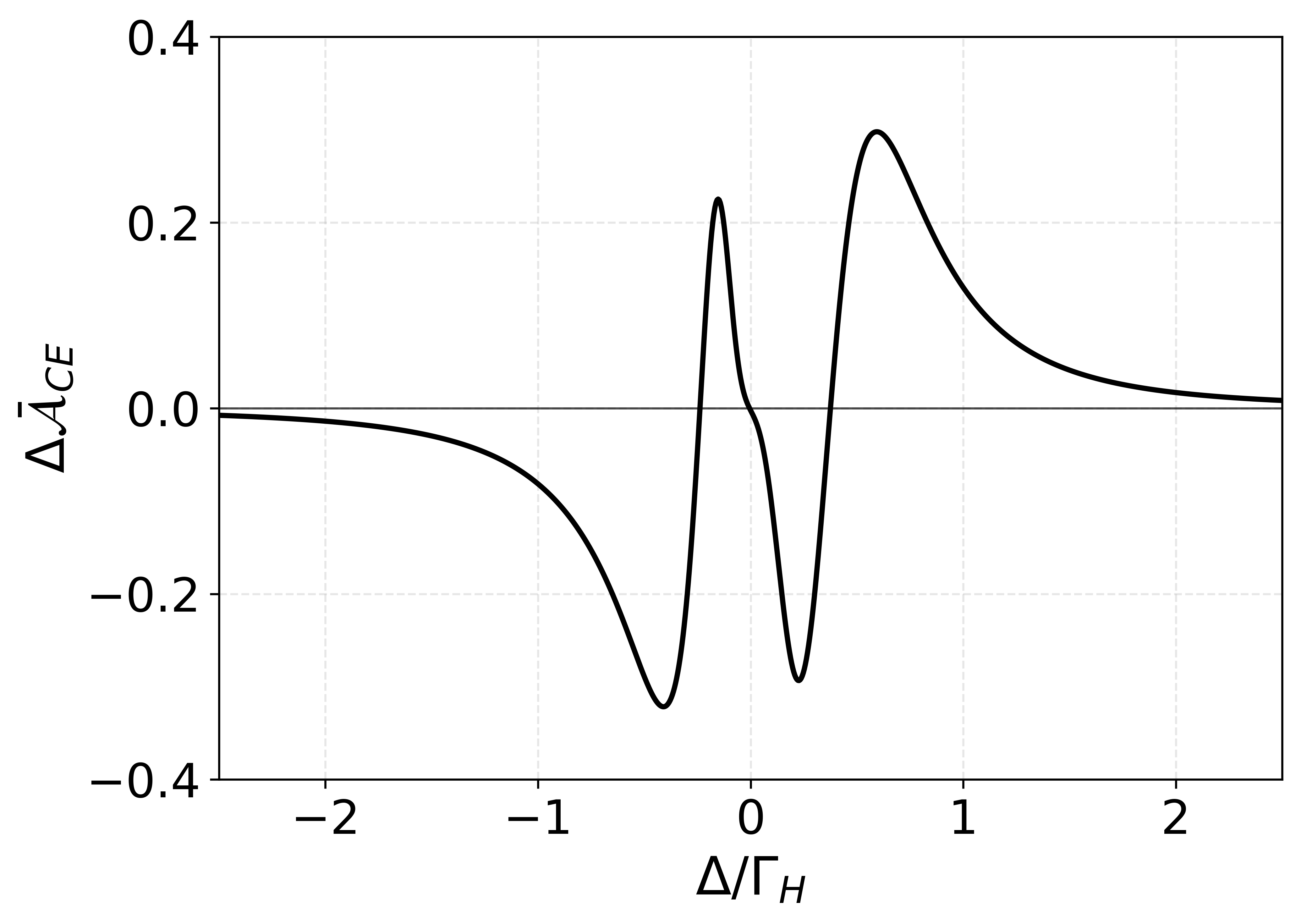}
		\caption{Effect of the non-Abelian hybrid-energy correction on the normalized CE response, shown as the difference $\Delta\mathcal{A}_{\text{CE}}(\Delta) = \tilde{\mathcal{A}}_{\text{CE}}^{\text{self}}(\Delta) - \tilde{\mathcal{A}}_{\text{CE}}^{0}(\Delta)$ against $\Delta/\Gamma_H$, where $\tilde{\mathcal{A}}_{\text{CE}}^{\text{self}}$ denotes the full model and $\tilde{\mathcal{A}}_{\text{CE}}^{0}$ the control case without gluon self-interaction. The pronounced negative feature near $\Delta = 0$ shows that the non-Abelian correction redistributes the absorptive response near resonance, shifting the transparency feature without destroying coherent interference.}
		\label{fig:self_interaction_effect}
	\end{figure}
	
	The specific role of gluon self-interaction is shown in Figure~\ref{fig:self_interaction_effect}, which plots the difference profile $\Delta\mathcal{A}_{\mathrm{CE}}(\Delta) = \tilde{\mathcal{A}}_{\mathrm{CE}}^{\mathrm{self}}(\Delta) - \tilde{\mathcal{A}}_{\mathrm{CE}}^{0}(\Delta)$ between the full model and the control case without self-interaction. The resulting curve maps the spectral redistribution induced by the non-Abelian hybrid-energy shift $\Delta\omega_H^{\text{NA}}(s)$. Near exact resonance ($\Delta \approx 0$), the difference shows a pronounced negative dip, while away from resonance it develops a small positive shoulder. This behavior shows that the non-Abelian correction redistributes the absorptive response across the spectral band rather than uniformly reducing it at all detunings. Consequently, in the present framework, the non-Abelian contribution mainly shifts the spectrum and modifies the line shape rather than generating the interference mechanism itself. While it sets the transparency feature’s spectral position, symmetry, and local depth, it does not determine whether dark-state interference exists. The EIT-like destructive interference dip is ensured by the topology of the $\Lambda$-type coupling scheme and low lower-state decoherence ($\gamma_{12} \ll \Gamma_H$).
	
\begin{figure}[ht]
		\centering
		\includegraphics[width=0.7\linewidth]{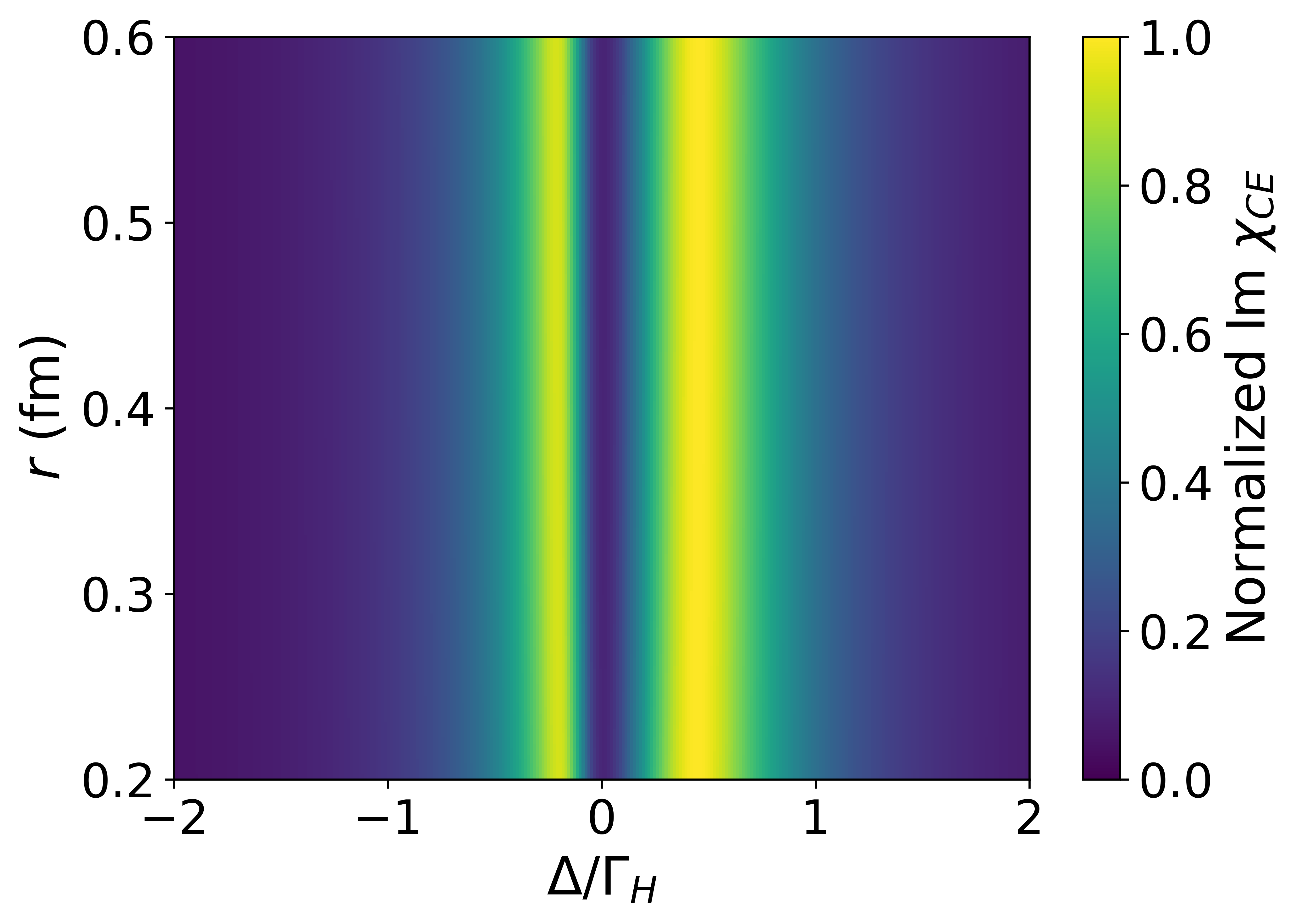}
		\caption{Globally normalized two-dimensional contour map of the CE response $\tilde{\mathcal{A}}_{\text{CE}}(r, \Delta)$ in the interquark separation versus detuning $(\Delta/\Gamma_H)$ plane for the full CE model at fixed flux-tube radius $s = 0.01\text{ fm}$. The dark horizontal band around $\Delta = 0$ marks an EIT-like transparency region where hybrid-channel excitation is suppressed by destructive interference. Its persistence along $r$ indicates that the dark-state spectral position is governed primarily by the two-photon resonance condition, while the flux-tube geometry mainly affects the amplitude and line shape of the sidebands.}
		\label{fig:detuning_distance_map}
		
	\end{figure}

	Figure~\ref{fig:detuning_distance_map} shows the dependence of the CE response on interquark separation $r$ as the two-dimensional contour map $\tilde{\mathcal{A}}_{\mathrm{CE}}(r,\Delta)$. The map is globally normalized over the $(r,\Delta)$ domain, allowing spatial and spectral variations to be evaluated on a unified scale. Its dominant feature is a narrow dark vertical stripe centered at $\Delta=0$, uninterrupted across the full range $r \in [0.2, 0.6]~\mathrm{fm}$. This low-response band marks the detuning interval where dark-state interference suppresses excitation of the hybrid channel. Around the central minimum, the bright, high-intensity regions indicate where the resonant envelope dominates the suppression. %The nearly uniform spectral width and position of this transparency band across different $r$ values show that, for a fixed flux-tube radius $s=0.010~\mathrm{fm}$, the EIT-like window is determined mainly by the two-photon resonance condition, while the flux-tube geometry primarily affects the overall response amplitude and sideband structure.
	\begin{figure}[ht]
		\centering
		\includegraphics[width=0.7\linewidth]{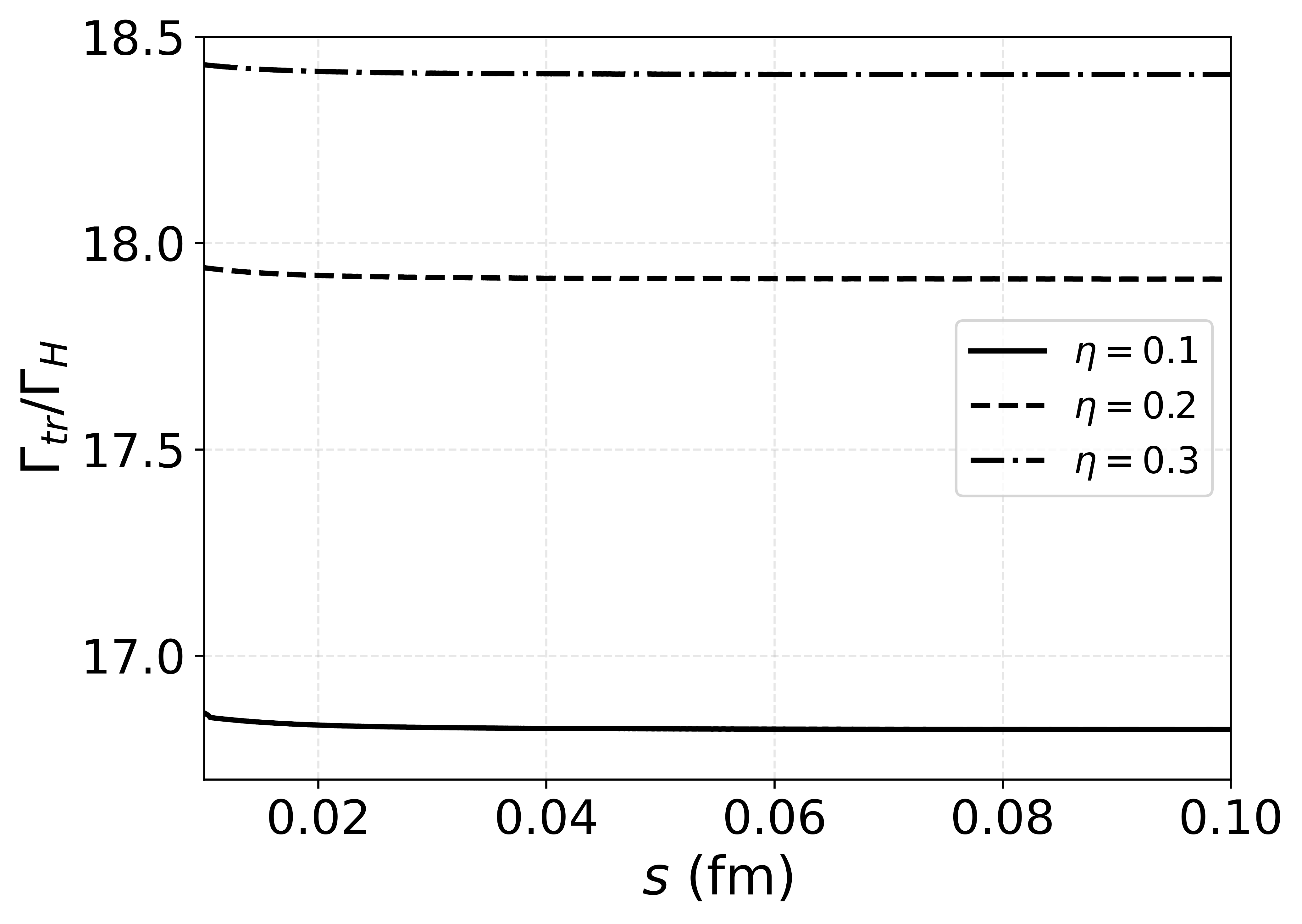}
		\caption{Normalized transparency-window width $\Gamma_{\text{tr}}/\Gamma_H$ as a function of flux-tube radius $s$ (in fm) for three thresholds: $\eta = 0.1$, $\eta = 0.2$, and $\eta = 0.3$. For all thresholds, $\Gamma_{\text{tr}}$ decreases monotonically as $s$ increases. This reflects the $s^{-2}$ dependence of the non-Abelian gluon self-interaction, equation~\eqref{eq:full_nonabelian_shift}, where narrower flux tubes (smaller $s$) strengthen the effective coupling and dark-state suppression, while wider flux tubes weaken this contribution and narrow the detuning range satisfying the transparency criterion.}
		\label{fig:width_vs_s}
	\end{figure}
	
	Finally, Figure~\ref{fig:width_vs_s} shows how the transparency-window width $\Gamma_{\mathrm{tr}}$ depends on the flux-tube radius $s$. For all evaluated thresholds ($\eta = 0.1, 0.2$, and $0.3$), $\Gamma_{\mathrm{tr}}$ decreases weakly and monotonically over the range of \(s\) shown. This behavior reflects the $s^{-2}$ scaling of the non-Abelian gluon self-interaction in equation~\eqref{eq:full_nonabelian_shift}: narrower flux tubes (smaller $s$) concentrate field energy, enhancing the effective CE coupling and reinforcing dark-state suppression. Broader flux tubes weaken this interaction, narrowing the detuning range over which the transparency criterion is satisfied. The threshold comparison also shows that while the quantitative magnitude of $\Gamma_{\mathrm{tr}}$ predictably expands under less restrictive criteria (larger $\eta$), its dependence on geometry remains stable. The monotonic decrease of $\Gamma_{\mathrm{tr}}$ with increasing $s$ is therefore a structural property of the model, not an artifact of the threshold parameter.
	
	%%%%%%%%%%%%%%%%%%%%%%%%%%%%%%%%%%%%%%%%%%%%%%%%%%%%
	\section{Dynamical window and observability of the confined-state transparency}\label{sec:time}
	
	%Of course, the mere existence of a dark-state in the reduced Hilbert space does not imply a measurable transparency signal. In fact, in a time-dependent realization, we separate three time scales. First, of course, is that the system requires a finite formation time to develop the coherent superposition associated with the dark-state. After that, the transparency response must persist long enough for the probe to register a measurable change in transmission or absorption. And at the end, the reduced description must remain valid during that interval, without significant leakage into configurations outside the effective three-state subspace.
	
	The relevant formation time is determined by the coherent couplings in the effective Hamiltonian \cite{ElectromagneticallyInducedTransparency.RevModPhys.77.633.2005}. Denoting by
	\begin{equation}\label{eq:time_1}
		\Omega_{\mathrm{eff}}\equiv 
		\sqrt{|\Omega_{1H}|^2+|\Omega_{2H}|^2},
	\end{equation} 
	this effective Rabi frequency combines the two coherent couplings into a single net coupling strength. Since the dark-state coherence is established by coherent mixing at this rate, the characteristic formation time is parametrically of order
	\begin{equation}\label{eq:time_2}
		\tau_{\mathrm{form}} \sim \Omega_{\mathrm{eff}}^{-1}.
	\end{equation}
	
	%The estimate \eqref{eq:time_2} indicates the minimum time for the off-diagonal density-matrix elements to become appreciable, and, of course, the transparency signal can be meaningfully defined only if the relevant coherent regime lasts longer than this formation time, given by
	%\begin{equation}
	%\tau_{\mathrm{avail}} \gg \tau_{\mathrm{form}},
	%\end{equation}
	%where \(\tau_{\mathrm{avail}}\) denotes the period during which the external parameters and the effective confined-state description remain approximately unchanged. 
	The available time is limited by several independent processes, and, in particular, the hybrid-state width introduces the relaxation time
	\begin{equation}
		\tau_H\sim\Gamma_H^{-1},
	\end{equation}
	while decoherence in the lower-state sector gives
	\begin{equation}
		\tau_{12}\sim\gamma_{12}^{-1}.
	\end{equation}
	
	It is important to stress that states omitted from the reduced Hilbert space can induce a leakage rate \(\Gamma_{\mathrm{leak}}\), and the surrounding medium can add a timescale \(\tau_{\mathrm{env}}\). Then, a conservative estimate of the effective coherent regime is given by the available time
	\begin{equation}
		\tau_{\mathrm{avail}} \sim \min\left(
		\Gamma_H^{-1}, \gamma_{12}^{-1}, \Gamma_{\mathrm{leak}}^{-1}, \tau_{\mathrm{env}}
		\right),
	\end{equation}
	resulting in the requirement for a dynamically resolvable transparency window is given by 
	\begin{equation}\label{eq:ineq_time}
		\Omega_{\mathrm{eff}}^{-1} \ll \tau_{\mathrm{avail}}.
	\end{equation}
	
	If inequality \eqref{eq:ineq_time} is not satisfied, then the dark-state coherence does not have time to form before the system relaxes, evolving away from the assumed configuration, or leaves the reduced subspace. This distinction is important because the lifetime of the intermediate hybrid configuration and the dark-state response are not necessarily identical. The dark state contains only a small admixture of the lossy hybrid component. In the idealized limit, the hybrid population vanishes at exact two-photon resonance, and decay of the intermediate state primarily determines the width and relaxation dynamics of the transparency feature rather than its full lifetime. However, the coherence between the two lower states remains vulnerable to the dephasing rate \(\gamma_{12}\). Thus, a long-lived dark-state transparency requires both sufficiently strong coherent coupling and a lower-state decoherence rate small enough to preserve the relevant superposition.
	
	%Considering the above comments, one has the following hierarchy
	%\begin{equation}
	%\Gamma_H^{-1}
	%\lesssim
	%\Omega_{\mathrm{eff}}^{-1}
	%\ll
	%\gamma_{12}^{-1},
	%\end{equation}
	%with
	
	However, it is important to note that a time-dependent treatment would make the lifetime criterion explicit. Starting from an initial density matrix \(\rho(0)\), one could evolve \cite{breuer_book,lindblad.1976}
	\begin{equation}
		\dot{\rho}(t)
		=
		-i\left[H_{\mathrm{eff}}(t),\rho(t)\right]
		+\mathcal{L}_{H}\!\left[\rho(t)\right]
		+\mathcal{L}_{12}\!\left[\rho(t)\right]
		+\mathcal{L}_{\mathrm{leak}}\!\left[\rho(t)\right].
	\end{equation}
	and using the instantaneous dark-state population,
	\begin{equation}
		P_D(t) =\langle D_{\mathrm{CE}}|\rho(t)|D_{\mathrm{CE}}\rangle,
	\end{equation}
	then the time-dependent probe susceptibility can be used to define an operational transparency lifetime. For example, one may define \(\tau_{\mathrm{EIT}}\) as the interval during which the hybrid-state population remains below a chosen threshold, or during which the absorptive part of the susceptibility stays below a specified fraction of its two-level value. In this sense, the dark-state population \(P_D(t)\) provides a natural population-based proxy for the same finite-time window, while the susceptibility criterion offers an operational measure of when the associated response remains observable. This definition would distinguish the formal existence of an eigenstate from the finite-time observability of its associated response.
	
	In practice, however, if the transparency lasts for a time \(\tau_{\mathrm{EIT}}\) shorter than the detector response time \(\tau_{\mathrm{det}}\), the effect is effectively averaged out. Then, an additional observability condition is
	\begin{equation}
		\tau_{\mathrm{EIT}} \gtrsim \tau_{\mathrm{det}},
	\end{equation}
	or, for a frequency-domain measurement, the associated spectral width must exceed the inverse observation time,
	\begin{equation}
		\Delta E_{\mathrm{EIT}} \gtrsim \tau_{\mathrm{obs}}^{-1},
	\end{equation}
	providing a concrete interpretation of the limited time scales associated with the three-state truncation. 
	
	Considering the above statements, the main result here is given by
	\begin{equation}
		\tau_{\mathrm{form}}
		\ll
		\tau_{\mathrm{EIT}}
		\lesssim
		\min\left(
		\Gamma_H^{-1},
		\gamma_{12}^{-1},
		\Gamma_{\mathrm{leak}}^{-1},
		\tau_{\mathrm{env}}
		\right),
		\qquad
		\tau_{\mathrm{EIT}}\gtrsim\tau_{\mathrm{det}}.
	\end{equation}
	taking into account that \(\Gamma_H^{-1}\) does not necessarily determine the dark-state lifetime since the population in the hybrid state is suppressed by interference. However, \(\Gamma_H^{-1}\) determines the relaxation scale and the width response scale. For the benchmark parameters, the relevant scales are
	\begin{equation}
		\Gamma_H^{-1}\simeq0.20\mathrm{fm}/c,\qquad
		\Omega_{\mathrm{eff}}^{-1}\simeq0.33\mathrm{fm}/c,\qquad
		\gamma_{12}^{-1}\simeq9.9~\mathrm{fm}/c,
	\end{equation}
	suggesting that a potentially measurable window exists only if the confined configuration and coupling parameters remain approximately constant for at least a few times $0.3~\mathrm{fm}/c$. Nevertheless, if the configuration lifetime is comparable to or shorter than this value, the stationary prediction should be understood only as a formal mechanism, not as an observable sign of transparency.
	
	%The physically relevant condition is thus not merely that a dark-state exists, but that there is a nonempty dynamical window satisfying
	%\begin{equation}
	%\tau_{\mathrm{form}} \ll \tau_{\mathrm{meas}}
	%\lesssim \tau_{\mathrm{avail}},
	%\end{equation}
	%with \(\tau_{\mathrm{meas}}\) also compatible with the temporal resolution of the measurement. 
	
	%Thus, the coherent dynamics can develop before lower-state dephasing becomes important, although the formation time is not parametrically shorter than the decay time of the hybrid component. This comparison is essential: a nominally narrow or strongly suppressed stationary absorption coefficient is insufficient if the coherent interaction lasts for less than one formation time or involves too few coherent cycles.
	
	%Considering the parameter values, the formation scale is 
	%\begin{equation}
	%\tau_{\mathrm{form}}\sim 0.3~\mathrm{fm}/c,    
	%\end{equation}
	%while the decoherence lifetime is 
	%\begin{equation}
	%\gamma_{12}^{-1}\sim 10~\mathrm{fm}/c,
	%\end{equation}
	%suggesting that a potentially measurable window exists only if the confined configuration and coupling parameters remain approximately constant for at least a few times $0.3~\mathrm{fm}/c$. Nevertheless, if the configuration lifetime is comparable to or shorter than this value, the stationary prediction should be understood only as a formal mechanism, not as an observable sign of transparency.
	
	%%%%%%%%%%%%%%%%%%%%%%%%%%%%%%%%%%%%%%%%%%%%%%%%%%%%%%%
	\section{Final remarks and outlook}
	\label{sec:final_remarks}
	
	In this work, we developed a minimal QCD-inspired model to study dark-state interference in a reduced subspace of confined color configurations. By mapping a static CE flux-tube background and time-dependent CE perturbations onto a $\Lambda$-type three-level scheme, we described coherent transitions between two low-energy $Q\bar{Q}$ states ($\lvert 1\rangle$, $\lvert 2\rangle$) and a gluon-excited hybrid state ($\lvert H\rangle$). Under two-photon resonance and Markovian GKSL dynamics, the system supports a dark state that decouples from the hybrid channel, suppressing the steady-state probe susceptibility $\mathrm{Im}\chi_{\mathrm{CE}}$ through destructive quantum interference. Although the model uses normalized parameters, lattice QCD calculations for realistic charmonium ($c\bar{c}$) systems place the lowest hybrid state about $1.0 - 1.5\text{ GeV}$ above the ground state \cite{dudek2008}. The dynamical discussion shows that the effect is meaningful only within a finite window set by coherent formation, decoherence, leakage, and environmental evolution. %In these systems, the decay width $\Gamma_H$ sets the timescale, and the transparency window remains accessible if the CE-induced transition rates exceed the medium's decoherence rate $\gamma_{12}$.
	
	The phenomenological non-Abelian correction modifies the position and shape of the transparency feature without generating the underlying interference. The response also retains a clear dependence on the flux-tube geometry. As shown in the response map $\tilde{\mathcal{A}}_{\mathrm{CE}}(r,\Delta)$, this transparency feature remains along the interquark coordinate $r$, changing the overall amplitude rather than the spectral position. Moreover, the operational transparency width $\Gamma_{\mathrm{tr}}$ decreases monotonically as the flux-tube radius $s$ increases, inheriting the $s^{-2}$ dependence of the non-Abelian self-interaction.
	
	The transparency window depends strongly on the medium's thermal evolution. In an open-quantum-system treatment, the lower-state decoherence rate $\gamma_{12}$ reflects QGP transport properties, including the jet-quenching parameter $\hat{q}$ and thermal dephasing rates. Well above $T_c$, strong collisional momentum broadening makes $\gamma_{12} \gg \lvert\Omega_{2H}\rvert^2/\Gamma_H$, destroys two-photon coherence, and restores a standard Lorentzian absorption profile. Near $T_c$, non-perturbative flux-tube remnants and dual magnetic monopoles can temporarily stabilize effective CE fields, allowing a low-decoherence regime ($\gamma_{12} \ll \Gamma_H$) to open a transient transparency window. Although thermal lattice QCD indicates that the free-energy string tension $\sigma_F(T)$ vanishes at $T_c$ while the internal-energy tension $\sigma_V(T)$ persists into the deconfined phase \cite{kaczmarek.2005,shuryak.2018}, the present dynamic approach instead treats thermal dissolution as temperature-dependent dephasing ($\gamma_{12}$) rather than a static change in string tension. The depth and spectral width $\Gamma_{\mathrm{tr}}$ of this window therefore provide an indication of the onset of quantum coherence loss in hadronic media.
	
	Possible phenomenological implications for heavy-quark observables may be investigated within future open-system extensions, where the transparency width could be related to quarkonium suppression observables in a qualitative manner, but no quantitative prediction for quarkonium yields or nuclear-modification factors is provided here.
	
	Several directions extend this framework. For example, the phenomenological parameterization used here can be constrained by matching energies and transition matrix elements to lattice QCD potential curves or short-distance Born-Oppenheimer approximations for heavy-quark hybrids. Of course, although gluonic self-interaction was treated as a diagonal energy shift, strong-coupling fluctuations may induce off-diagonal mixing between states $\lvert 1\rangle$ and $\lvert 2\rangle$. Moreover, the Markovian approximation in the GKSL equation can be refined by adding non-Markovian memory kernels to capture short-time gluonic fluctuations and memory effects in high-energy collisions. Finally, extending the $\Lambda$-scheme to $N$-level configurations could reveal multi-frequency dark resonances and interference patterns in hadronic spectra. 
	
	Because the coherent dynamics are too fast for direct time resolution, the effect would have to be inferred from time-integrated observables such as decay yields, spectral line shapes, channel ratios, or final-state correlations. Such an interpretation would require a statistically significant deviation from predictions by incoherent or reduced models.
	
	While atomic-physics analogies have clear limitations (quarks and gluons do not form literal atomic orbitals), the present results show that confined color configurations coupled to a common gluon-excited channel can combine into a coherent superposition that dynamically decouples from that channel. The key physics result is that the hadronic mapping offers a controlled, quantitative framework for capturing confinement-driven quantum interference structures in strongly interacting matter, highlighting its role as a modeling tool rather than a literal atomic analogy.

	\section*{Acknowledgments}
	
	SDC acknowledges the Federal University of São Carlos for its institutional support.
	
	%%%%%%%%%%%%%%%%%%%%%%%%%%%%%%%%%%%%%%%%%%%%%%%
	\appendix
	%%%%%%%%%%%%%%%%%%%%%%%%%%%5
	\section{Derivation of the reduced CE interaction Hamiltonian}
	\label{app:reduced_interaction}
	
	Here, we obtain the reduced interaction Hamiltonian used in
	equation~\eqref{eq:interaction_reduced_basis} from the effective CE transition operator was introduced by the equation~\eqref{eq:chromodipole_interaction}. We begin with equation~\eqref{eq:chromodipole_interaction}
	%\begin{equation}
	%\hat{V}(t) = -g\,\hat{\mathbf{d}}_{\mathrm{eff}}^{\,a}
	%\cdot\delta\boldsymbol{\mathcal{E}}^{a}(r,s,t),
	%\label{eq:appendix_chromodipole_interaction}
	%\end{equation}
	where $\hat{\mathbf{d}}_{\mathrm{eff}}^{\,a}$ acts within the reduced
	Hilbert space given by~\eqref{eq:effective_hilbert_space}. 
	%\begin{equation}
	%\mathcal{H}_{\mathrm{eff}} =
	%\operatorname{span}
	%\left\{
	%\lvert 1\rangle,
	%\lvert 2\rangle,
	%\lvert H\rangle
	%\right\}.
	%\end{equation}
	The operator should be understood as the projection of the underlying
	QCD CE interaction onto the truncated confined-state subspace.
	
	%%%%%%%%%%%%%%%%%%%%%%%%%
	\subsection{Two-component CE driving field}
	
	A $\Lambda$-type interference mechanism requires two coherent transition amplitudes that connect the two lower configurations to the same hybrid configuration. We therefore decompose the time-dependent CE perturbation into probe-like and coupling-like components,
	\begin{equation}
		\delta\boldsymbol{\mathcal{E}}^{a}(r,s,t) =
		\boldsymbol{\mathcal{E}}_{p}^{a}(r,s) \cos(\omega_{p}t)
		+ \boldsymbol{\mathcal{E}}_{c}^{a}(r,s)
		\cos(\omega_{c}t),
		\label{eq:appendix_two_component_field}
	\end{equation}
	where $\omega_{p}$ and $\omega_{c}$ are the corresponding driving
	frequencies. Using
	\begin{equation}
		\cos(\omega t) = \frac{1}{2} \left(
		e^{-i\omega t} + e^{i\omega t} \right),
		\label{eq:appendix_cosine_decomposition}
	\end{equation}
	equation~\eqref{eq:appendix_two_component_field} becomes
	\begin{equation}
		\begin{split}
			\delta\boldsymbol{\mathcal{E}}^{a}(r,s,t) = \frac{1}{2}
			\Big[
			& \boldsymbol{\mathcal{E}}_{p}^{a}(r,s)
			\left( e^{-i\omega_{p}t} + e^{i\omega_{p}t} \right)
			+  \boldsymbol{\mathcal{E}}_{c}^{a}(r,s) \left(e^{-i\omega_{c}t}
			+ e^{i\omega_{c}t} \right) \Big].
		\end{split}
		\label{eq:appendix_field_frequency_decomposition}
	\end{equation}
	
	Substitution into equation~\eqref{eq:chromodipole_interaction} %{eq:appendix_chromodipole_interaction}
	gives
	\begin{equation}
		\begin{split}
			\hat{V}(t) = -\frac{g}{2} \hat{\mathbf{d}}_{\mathrm{eff}}^{\,a}
			\cdot\Big[& \boldsymbol{\mathcal{E}}_{p}^{a}(r,s)
			\left(e^{-i\omega_{p}t}+e^{i\omega_{p}t}\right)+ \boldsymbol{\mathcal{E}}_{c}^{a}(r,s)\left(
			e^{-i\omega_{c}t}+e^{i\omega_{c}t}\right)\Big].
		\end{split}
		\label{eq:appendix_interaction_frequency_decomposition}
	\end{equation}
	
	%%%%%%%%%%%%%%%%%%%%%%%%%%%%%
	\subsection{Projection onto the reduced basis}
	
	Within $\mathcal{H}_{\mathrm{eff}}$, the transition operator can be
	expanded as
	\begin{equation}
		\hat{\mathbf{d}}_{\mathrm{eff}}^{\,a} = \sum_{i,j\in\{1,2,H\}} \mathbf{d}_{ij}^{a} \lvert i\rangle\langle j\rvert,
		\label{eq:appendix_transition_operator_expansion}
	\end{equation}
	where $\mathbf{d}_{ij}^{a}$ obeys the relation~\eqref{eq:ortho_operator}.
	%\begin{equation}
	%\mathbf{d}_{ij}^{a} = \langle i|
	%\hat{\mathbf{d}}_{\mathrm{eff}}^{\,a} |j\rangle.
	%\label{eq:appendix_transition_matrix_elements}
	%\end{equation}
	
	For the $\Lambda$-type configuration considered in this work, we retain only the matrix elements that connect the two lower confined configurations to the common hybrid configuration. Thus,
	\begin{equation}
		\begin{split}
			\hat{\mathbf{d}}_{\mathrm{eff}}^{\,a} \simeq
			\mathbf{d}_{H1}^{a} \lvert H\rangle\langle 1\rvert
			+ \mathbf{d}_{H2}^{a} \lvert H\rangle\langle 2\rvert
			+ \mathbf{d}_{1H}^{a} \lvert 1\rangle\langle H\rvert
			+ \mathbf{d}_{2H}^{a} \lvert 2\rangle\langle H\rvert.
		\end{split}
		\label{eq:appendix_lambda_operator}
	\end{equation}
	
	Hermiticity requires
	\begin{equation}
		\mathbf{d}_{1H}^{a} = \left( \mathbf{d}_{H1}^{a} \right)^{\ast},
		\qquad
		\mathbf{d}_{2H}^{a} = \left( \mathbf{d}_{H2}^{a} \right)^{\ast}.
		\label{eq:appendix_hermiticity}
	\end{equation}
	
	We further assume that the probe-like component predominantly couples
	the transition
	\begin{equation}
		\lvert 1\rangle \longleftrightarrow \lvert H\rangle,
	\end{equation}
	whereas the coupling-like component predominantly couples
	\begin{equation}
		\lvert 2\rangle \longleftrightarrow \lvert H\rangle.
	\end{equation}
	
	Cross-couplings are neglected in the minimal model. Including them would add off-resonant terms but would not change the basic derivation presented below.
	
	Substituting equation~\eqref{eq:appendix_lambda_operator} into
	equation~\eqref{eq:appendix_interaction_frequency_decomposition}, and
	retaining only the intended pairings between field components and
	transitions, gives
	\begin{equation}
		\begin{split}
			\hat{V}(t) \simeq -\frac{g}{2} \Big[ & \left(
			\mathbf{d}_{H1}^{a} \cdot \boldsymbol{\mathcal{E}}_{p}^{a}
			\right) \left(e^{-i\omega_{p}t}+e^{i\omega_{p}t}\right)
			\lvert H\rangle\langle 1\rvert
			+  \left(\mathbf{d}_{H2}^{a}\cdot\boldsymbol{\mathcal{E}}_{c}^{a}
			\right)\left(e^{-i\omega_{c}t}+e^{i\omega_{c}t}\right)
			\lvert H\rangle\langle 2\rvert
			+ \mathrm{h.c.} \Big].
		\end{split}
		\label{eq:appendix_projected_interaction}
	\end{equation}
	
	%%%%%%%%%%%%%%%%%%%%%%%%%%%%%%%%
	\subsection{Interaction picture and rotating-wave approximation}
	
	The slowly varying terms are most easily identified in the interaction
	picture defined by the unperturbed Hamiltonian
	\begin{equation}
		\hat{H}_{0} = \omega_{1}\lvert 1\rangle\langle 1\rvert
		+\omega_{2}\lvert 2\rangle\langle 2\rvert
		+\omega_{H}\lvert H\rangle\langle H\rvert.
		\label{eq:appendix_unperturbed_hamiltonian}
	\end{equation}
	
	An operator $\lvert i\rangle\langle j\rvert$ evolves according to
	\begin{equation}
		e^{i\hat{H}_{0}t} \lvert i\rangle\langle j\rvert
		e^{-i\hat{H}_{0}t} = e^{i(\omega_{i}-\omega_{j})t}
		\lvert i\rangle\langle j\rvert.
		\label{eq:appendix_operator_time_evolution}
	\end{equation}
	
	Therefore,
	\begin{equation}
		\lvert H\rangle\langle 1\rvert \longrightarrow
		e^{i(\omega_{H}-\omega_{1})t} \lvert H\rangle\langle 1\rvert,\qquad
		\label{eq:appendix_H1_phase}
		%\end{equation}
		%and
		%\begin{equation}
		\lvert H\rangle\langle 2\rvert \longrightarrow e^{i(\omega_{H}-\omega_{2})t}
		\lvert H\rangle\langle 2\rvert.
		%\label{eq:appendix_H2_phase}
	\end{equation}
	
	The probe-like contribution then contains the phase factors
	\begin{equation}
		e^{i(\omega_{H}-\omega_{1}-\omega_{p})t},
		\qquad
		e^{i(\omega_{H}-\omega_{1}+\omega_{p})t},
		\label{eq:appendix_probe_phases}
	\end{equation}
	whereas the coupling-like contribution contains
	\begin{equation}
		e^{i(\omega_{H}-\omega_{2}-\omega_{c})t},
		\qquad
		e^{i(\omega_{H}-\omega_{2}+\omega_{c})t}.
		\label{eq:appendix_coupling_phases}
	\end{equation}
	
	Close to resonance, $\omega_{p} \simeq \omega_{H}-\omega_{1}$, and 
	$\omega_{c} \simeq \omega_{H}-\omega_{2}$, the terms proportional to
	\begin{equation}
		e^{i(\omega_{H}-\omega_{1}-\omega_{p})t}, \qquad
		%\end{equation}
		%and
		%\begin{equation}
		e^{i(\omega_{H}-\omega_{2}-\omega_{c})t}
	\end{equation}
	vary slowly. By contrast, the counter-rotating terms contain sums of the driving and transition frequencies and oscillate rapidly. The rotating-wave approximation neglects these rapidly oscillating terms. Thus, after the rotating-wave approximation, the interaction Hamiltonian becomes
	\begin{equation}
		\begin{split}
			\hat{V}_{\mathrm{RWA}}(t) = -\frac{g}{2}
			\Big[&\left(\mathbf{d}_{H1}^{a}\cdot\boldsymbol{\mathcal{E}}_{p}^{a}(r,s)
			\right)e^{-i\omega_{p}t}\lvert H\rangle\langle 1\rvert
			+\left(\mathbf{d}_{H2}^{a}\cdot\boldsymbol{\mathcal{E}}_{c}^{a}(r,s)
			\right)e^{-i\omega_{c}t}\lvert H\rangle\langle 2\rvert + \mathrm{h.c.}
			\Big].
		\end{split}
		\label{eq:appendix_interaction_rwa_pre_rabi}
	\end{equation}
	
	%%%%%%%%%%%%%%%%%%%%%%%%%%%%%%%%%%%%%%%%%
	\subsection{Definition of the effective Rabi frequencies}
	
	We define the effective CE Rabi frequencies as
	\begin{align}
		\Omega_{1H}(r,s) = -g\,
		\mathbf{d}_{H1}^{a} \cdot \boldsymbol{\mathcal{E}}_{p}^{a}(r,s),
		\qquad
		\Omega_{2H}(r,s) = -g\,\mathbf{d}_{H2}^{a}
		\cdot \boldsymbol{\mathcal{E}}_{c}^{a}(r,s).
		\label{eq:appendix_rabi_2H}
	\end{align}
	
	The factor $1/2$ in the interaction Hamiltonian is inherited from the decomposition of the real fields into positive- and negative-frequency components. With these definitions, equation~\eqref{eq:appendix_interaction_rwa_pre_rabi} reduces to
	\begin{equation}
		%\boxed{
			\hat{V}_{\mathrm{RWA}}(t) = \frac{1}{2}
			\left[\Omega_{1H}(r,s)e^{-i\omega_{p}t}\lvert H\rangle\langle 1\rvert
			+\Omega_{2H}(r,s)e^{-i\omega_{c}t}\lvert H\rangle\langle 2\rvert
			+\mathrm{h.c.}\right]
			%}
		\label{eq:appendix_interaction_reduced_basis}
	\end{equation}
	which is the reduced interaction Hamiltonian employed in
	equation~\eqref{eq:interaction_reduced_basis}. %The derivation can be summarized as
	%\begin{equation}
	%-g\, \hat{\mathbf{d}}_{\mathrm{eff}}^{\,a}
	%\cdot \delta\boldsymbol{\mathcal{E}}^{a}
	%\quad
	%\longrightarrow \quad \begin{array}{c}
		%\text{projection onto }
		%\{
		%\lvert 1\rangle,
		%\lvert 2\rangle,
		%\lvert H\rangle
		%\}
		%\\[0.2em]
		%+
		%\text{selection of the two } \Lambda\text{-type transitions}
		%\\[0.2em]
		%+
		%\text{frequency decomposition}
		%\\[0.2em]
		%+
		%\text{rotating-wave approximation}
		%\end{array}
		%\quad
		%\longrightarrow
		%\quad
		%\hat{V}_{\mathrm{RWA}}(t).
		%\end{equation}
		
		%%%%%%%%%%%%%%%%%%%%%%%%%%%%%%%%%%%%%%%%%%%%%%%%%%%
		\section{Derivation of the Effective Hamiltonian in the Rotating Frame}
		\label{app:RWA_derivation}
		
		In this Appendix, we present a step-by-step derivation of the effective Hamiltonian $\hat{H}_{\mathrm{eff}}$ in the reduced three-state basis $\mathcal{B} = \{\lvert 1\rangle, \lvert 2\rangle, \lvert H\rangle\}$, starting from the time-dependent physical Hamiltonian and using the Rotating-Wave Approximation (RWA).
		
		%%%%%%%%%%%%%%%%%%%%%%%%%%%%%
		\subsection{Unperturbed Hamiltonian and Interaction V(t)}
		
		Let the unperturbed energies of the confined low-energy states $\lvert 1\rangle$, $\lvert 2\rangle$, and the gluon-excited hybrid state $\lvert H\rangle$ be denoted by $E_1$, $E_2$, and $E_H$, respectively. Setting the energy scale such that $E_1 = 0$, the unperturbed Hamiltonian $\hat{H}_0$ in the basis $\mathcal{B}$ is written as
		\begin{equation}
			\hat{H}_{0} = \hbar 
			\begin{pmatrix}
				0 & 0 & 0 \\
				0 & \omega_{21} & 0 \\
				0 & 0 & \omega_{H1}
			\end{pmatrix},
			\label{eq:app_H0}
		\end{equation}
		where $\omega_{21} = (E_2 - E_1)/\hbar$ and $\omega_{H1} = (E_H - E_1)/\hbar$.
		
		The time-dependent interaction describing the coupling with the probe-like (frequency $\omega_p$) and coupling-like (frequency $\omega_c$) chromoelectric perturbations is given by equation~\eqref{eq:chromodipole_interaction}. Restricting the dipole transitions to the channels $\lvert 1\rangle \leftrightarrow \lvert H\rangle$ and $\lvert 2\rangle \leftrightarrow \lvert H\rangle$, the interaction Hamiltonian in matrix form takes the explicit time-dependent structure
		\begin{equation}
			\hat{V}(t) = \frac{\hbar}{2}
			\begin{pmatrix}
				0 & 0 & \Omega_{1H} \left(e^{i\omega_p t} + e^{-i\omega_p t}\right) \\
				0 & 0 & \Omega_{2H} \left(e^{i\omega_c t} + e^{-i\omega_c t}\right) \\
				\Omega_{1H}^* \left(e^{i\omega_p t} + e^{-i\omega_p t}\right) & \Omega_{2H}^* \left(e^{i\omega_c t} + e^{-i\omega_c t}\right) & 0
			\end{pmatrix},
			\label{eq:app_Vt}
		\end{equation}
		where $\Omega_{1H}$ and $\Omega_{2H}$ are the effective chromoelectric Rabi frequencies defined in equations~\eqref{eq:rabi_1H} and \eqref{eq:rabi_2H}.
		
		%%%%%%%%%%%%%%%%%%%%%%%%%%%%%%%%
		\subsection{Unitary Transformation to the Rotating Frame}
		
		To eliminate the fast explicit time dependencies, we transform the state vectors to a rotating frame via a time-dependent unitary operator $\hat{U}(t) = \exp(i\hat{S}t)$,
		\begin{equation}
			\lvert \psi_{\mathrm{rot}}(t)\rangle = \hat{U}(t) \lvert \psi(t)\rangle,
			\label{eq:app_unitary_transf}
		\end{equation}
		where $\hat{S}$ is a diagonal operator chosen to match the frequencies of the driving fields:
		\begin{equation}
			\hat{S} = 
			\begin{pmatrix}
				0 & 0 & 0 \\
				0 & \omega_p - \omega_c & 0 \\
				0 & 0 & \omega_p
			\end{pmatrix}.
			\label{eq:app_S_operator}
		\end{equation}
		
		Inserting equation~\eqref{eq:app_unitary_transf} into the time-dependent Schrödinger equation $i\hbar \frac{\partial}{\partial t}\lvert \psi(t)\rangle = (\hat{H}_0 + \hat{V}(t))\lvert \psi(t)\rangle$ yields the equation of motion in the rotating frame,
		\begin{equation}
			i\hbar \frac{\partial}{\partial t}\lvert \psi_{\mathrm{rot}}(t)\rangle = \hat{H}_{\mathrm{rot}}(t) \lvert \psi_{\mathrm{rot}}(t)\rangle,
		\end{equation}
		where the transformed Hamiltonian is given by
		\begin{equation}
			\hat{H}_{\mathrm{rot}}(t) = \hat{U}(t) \hat{H}_0 \hat{U}^\dagger(t) + \hat{U}(t) \hat{V}(t) \hat{U}^\dagger(t) - \hbar \hat{S}.
			\label{eq:app_Hrot_def}
		\end{equation}
		
		Evaluating the diagonal part $\hat{H}_0' = \hat{U}(t) \hat{H}_0 \hat{U}^\dagger(t) - \hbar \hat{S}$, we find:
		\begin{equation}
			\hat{H}_0' = \hbar
			\begin{pmatrix}
				0 & 0 & 0 \\
				0 & \omega_{21} - (\omega_p - \omega_c) & 0 \\
				0 & 0 & \omega_{H1} - \omega_p
			\end{pmatrix}.
			\label{eq:app_H0_prime}
		\end{equation}
		
		Defining the probe detuning from the hybrid state as $\Delta_H \equiv \omega_p-\omega_{H1}$ and the two-photon (Raman) detuning between the two lower states as $\delta \equiv \omega_{21} - (\omega_p - \omega_c)$, equation~\eqref{eq:app_H0_prime} simplifies to
		\begin{equation}
			\hat{H}_0' = \hbar
			\begin{pmatrix}
				0 & 0 & 0 \\
				0 & \delta & 0 \\
				0 & 0 & \Delta_H
			\end{pmatrix}.
		\end{equation}
		
		%%%%%%%%%%%%%%%%%%%%%%%%%%%%
		\subsection{Rotating-Wave Approximation}
		
		Next, we evaluate the interaction term in the rotating frame, $\hat{V}_{\mathrm{rot}}(t) = \hat{U}(t) \hat{V}(t) \hat{U}^\dagger(t)$. Using the explicit form of $\hat{U}(t) = \mathrm{diag}(1, e^{i(\omega_p-\omega_c)t}, e^{i\omega_p t})$, the off-diagonal elements become:
		\begin{align}
			\left[\hat{V}_{\mathrm{rot}}(t)\right]_{1H} &= \frac{\hbar}{2} \Omega_{1H} \left( 1 + e^{2i\omega_p t} \right), \\
			\left[\hat{V}_{\mathrm{rot}}(t)\right]_{2H} &= \frac{\hbar}{2} \Omega_{2H} \left( 1 + e^{2i\omega_c t} \right).
		\end{align}
		
		Under the standard RWA, which is valid when the optical driving frequencies $\omega_p$ and $\omega_c$ are much larger than the coupling strengths $|\Omega_{1H}|$, $|\Omega_{2H}|$, and the detunings, the rapidly oscillating terms $e^{\pm 2i\omega_p t}$ and $e^{\pm 2i\omega_c t}$ average out to zero on the physical timescales of the population dynamics. Neglecting these high-frequency terms, $\hat{V}_{\mathrm{rot}}$ becomes completely time-independent
		\begin{equation}
			\hat{V}_{\mathrm{RWA}} = \frac{\hbar}{2}
			\begin{pmatrix}
				0 & 0 & \Omega_{1H} \\
				0 & 0 & \Omega_{2H} \\
				\Omega_{1H}^* & \Omega_{2H}^* & 0
			\end{pmatrix}.
		\end{equation}
		
		Combining $\hat{H}_0'$ and $\hat{V}_{\mathrm{RWA}}$ (and setting $\hbar = 1$ in natural units), the total effective Hamiltonian $\hat{H}_{\mathrm{eff}} = \hat{H}_0' + \hat{V}_{\mathrm{RWA}}$ in the ordered basis $\{\lvert 1\rangle, \lvert 2\rangle, \lvert H\rangle\}$ takes the static matrix form:
		\begin{equation}
			\hat{H}_{\mathrm{eff}} = \frac{1}{2}
			\begin{pmatrix}
				0 & 0 & \Omega_{1H} \\
				0 & 2\delta & \Omega_{2H} \\
				\Omega_{1H}^{\ast} & \Omega_{2H}^{\ast} & 2\Delta_{H}
			\end{pmatrix},
		\end{equation}
		which coincides with equation~\eqref{eq:Heff_dark}.
		
		\bibliographystyle{unsrt}
		\bibliography{biblio}

	\end{document}